%% file: main_text.tex
\documentclass[
superscriptaddress,
amsmath,
amssymb,
achemso,
aps,
prl,
twocolumn,
]{revtex4-2}

\usepackage{braket}
\usepackage{graphicx}
\usepackage[colorlinks=true, allcolors=blue]{hyperref}
\usepackage{gensymb}
\usepackage{amsmath}
\usepackage{derivative}
\usepackage{upgreek}
\usepackage{float}  
\usepackage{xcolor}
\usepackage{footmisc}
\usepackage{siunitx}
\usepackage{calrsfs}
\usepackage{mathrsfs}
\usepackage{booktabs}
\usepackage{mathtools}

\newcommand{\ca}{$^{40}$Ca$^+$}

\newsavebox\mcFcontent
\savebox\mcFcontent{$\mathcal{F}$}

\makeatother

\begin{document}

\title{Super-resolution of two Closely-spaced Electromagnetic Fields via Walsh-Modulated Dynamical Decoupling Spectroscopy}

\author{Hao Wu}
\thanks{hao.wu@physics.ucla.edu}
\author{Grant D. Mitts}
\author{Clayton Z. C. Ho}
\author{Joshua A. Rabinowitz}
\author{Eric R. Hudson}
\affiliation{Department of Physics and Astronomy, University of California Los Angeles, Los Angeles, CA, USA}
\affiliation{Challenge Institute for Quantum Computation, University of California Los Angeles, Los Angeles, CA, USA}
\affiliation{Center for Quantum Science and Engineering, University of California Los Angeles, Los Angeles, CA, USA}
    
\date{\today}

\begin{abstract}

Due to quantum fluctuations, non-orthogonal quantum states cannot be distinguished with complete certainty, making their underlying physical parameters difficult to resolve.
Traditionally, it has been believed that the linewidth of a system behaves like these quantum fluctuations to set the ultimate limit on frequency resolution when two oscillating electromagnetic fields are applied.
Consequently, the measurement time $T$ required to resolve a frequency difference $\Delta \omega$ was assumed to diverge as $\Delta \omega \rightarrow 0$.
In this work, we show that linewidth does not play a defining role in resolving two closely spaced frequencies. 
Instead, the ultimate limit is set by parameter-independent quantum fluctuations, such as shot noise in our case.
We propose and experimentally demonstrate the first general broadband protocol for super-resolution spectroscopy.
Specifically, our protocol uses a Walsh-modulated dynamical decoupling (WMDD) sequence to encode the frequency separation $\Delta \omega$ between two unknown tones into a quantum state.
This leverages phase information to suppress parameter-independent shot noise, thereby enhancing the signal-to-noise ratio and enabling super-resolution spectroscopy. 
With this approach, we resolve two randomly chosen oscillating electric fields of order $10^{2} \textrm{ MHz}$ separated by $5 \textrm{ Hz}$, with a measured frequency difference of $5.0(1.6) \textrm{ Hz}$ using a measurement time per run of just $1 \textrm{ ms}$, representing an improvement of $200 \times$ beyond the traditional spectral resolution limit.
As such, our technique accelerates data acquisition by more than five orders of magnitude compared to conventional methods. 
Crucially, as our protocol is rooted in the motional Raman (quantum vector signal analyzer) framework, it is effective across an arbitrary frequency range and thus promises to enhance broadband sensing of electromagnetic fields and improve spectral efficiency of next-generation communication systems.
\end{abstract}

\maketitle

\section{Introduction}

The roots of modern quantum metrology can be traced to studies of the distinguishability of quantum states~\cite{Helstrom1969, Wootters1981}.
For two systems parametrized by $\theta$ and described by density operators $\hat{\rho}(\theta), \ \hat{\rho}(\theta + d\theta)$, a statistical distance $d = d\theta \sqrt{\mathcal{F}_Q{\left(\theta\right)}}$ can be defined~\cite{Braunstein1994}.
Intuitively, this statistical distance can be understood as roughly the number of distinguishable states separating two quantum states in a measurement.
Here, $\mathcal{F}_Q{\left(\theta\right)}$ is the quantum Fisher information (QFI), which sets a lower bound on the measured precision of $\theta$ in $N$ measurements of the system $\hat{\rho}\left(\theta \right)$ via the quantum Cramer-Rao bound (QCRB) $\sigma_\theta \geq 1/\sqrt{N\mathcal{F}_Q{\left(\theta\right)}}$. 
Current efforts aim to improve measurement precision by using quantum states that provide a larger $\mathcal{F}_Q$ and developing optimized protocols that allow the Fisher information $F_c$ of a given measurement to saturate the bound set by $\mathcal{F}_Q$~\cite{Degen2017,Pezze2018, Bouchet2021, Vittorio2004}.
To reach the QCRB, most work has relied on Rabi or Ramsey style protocols to tie $\theta$ to a quantity of interest, e.g. a magnetic field, and a post-dynamics optimization of the measurement basis, e.g. eigenstates of the symmetric logarithmic derivative \cite{Braunstein1994}.

Generally, two states become distinguishable once the statistical distance exceeds the underlying quantum fluctuations.
A fundamental metrological challenge thus arises when the statistical distance between two quantum states goes toward zero~\cite{Wootters1981}.
Such situations largely originate when quantum states resulting from the evolution due to two signals - e.g. closely spaced frequencies or spatial features - overlap substantially within the resolution limit (e.g. linewidth or diffraction limit) set by the standard measurement basis.
As a result, their distinguishability is reduced - $F_c$ vanishes - and the underlying parameters cannot be reliably measured. 
For decades, this phenomenon, known as Rayleigh’s curse~\cite{Tsang2016}, has been thought to set an ultimate limit in both frequency spectroscopy and imaging, as it was believed that the conventional resolution limit behaves equivalently to quantum fluctuations for discriminating quantum states.

Recent breakthroughs in super-resolution imaging, notably Ref.~\cite{Tsang2016}, have challenged this convention.
For two incoherent sources, $\mathcal{F}_Q$ is, in fact, nonzero, regardless of the traditional resolution limit, indicating that a non-vanishing $F_c$ is realizable.
Ref. ~\cite{Tsang2016} proposed that, by strategically tailoring the measurement basis, the previously neglected phase information within an electromagnetic field can be extracted, allowing $F_c$ to saturate $\mathcal{F}_Q$.
This has been experimentally demonstrated in Ref. ~\cite{Paur2016, Tang2016, Yang2016, Tham2017}.

Transitioning from spatial to frequency super-resolution is nontrivial - measurement techniques developed for imaging often lack clear analogues in frequency-sensing platforms.
To the best of our knowledge, the sole experimental demonstration of frequency super-resolution to date that does not map frequency onto spatial information was realized using nitrogen-vacancy (NV) centers in diamond~\cite{Rotem2019}.
It relied on time-correlated phase-sensitive measurements and a clever post-processing technique that averages over the phase and amplitude distributions of two closely spaced signals, achieving resolutions an order of magnitude below the conventional spectral limit.
Ref.~\cite{Gefen2019} further highlighted the importance of signal incoherence as a necessary condition for super-resolution and proposed a general framework for achieving super-resolution by engineering a measurement process that nulls quantum projection noise, along with an implementation for magnetic-field detection with NV-centers which relies on precise and well-timed control sequences.
\begin{figure*}
    \centering
    \includegraphics[width = 1\textwidth]{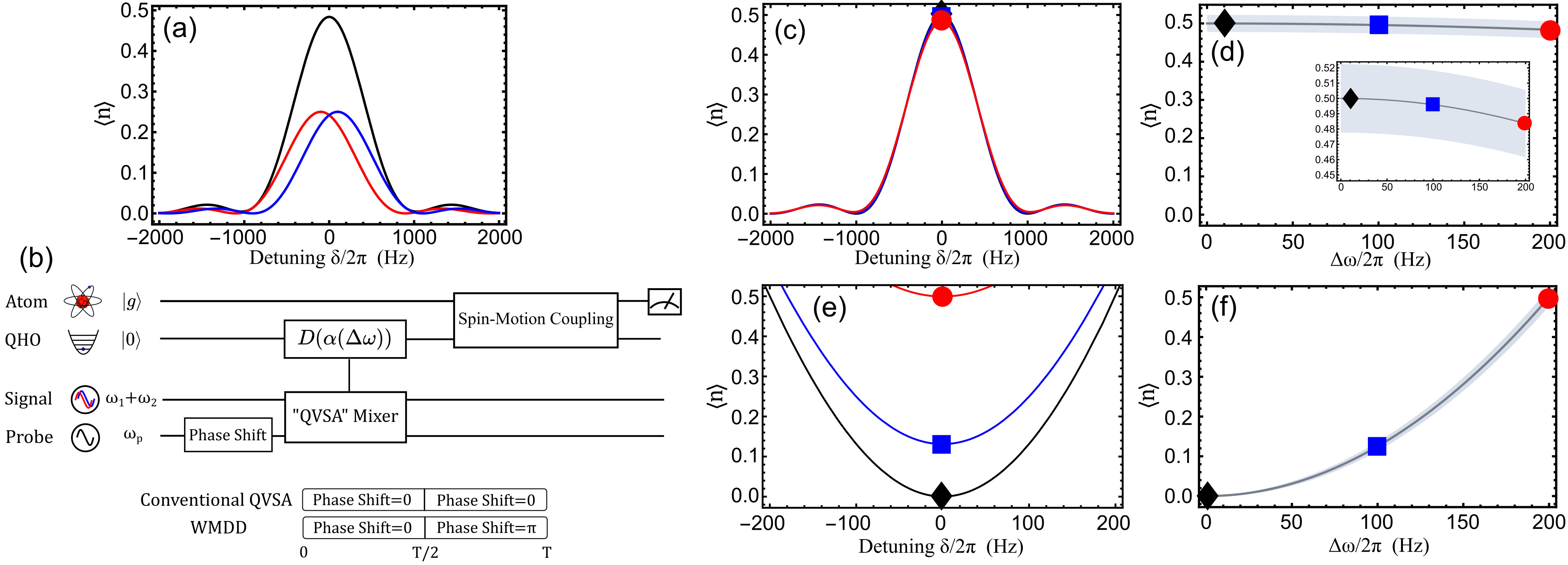}
    \caption{Spectroscopy of two incoherent tones using conventional quantum vector signal analysis (QVSA) and Walsh-modulated dynamic decoupling (WMDD) techniques.
    (a) Conventional spectroscopy with two incoherent signal tones.
    When the separation between the two tones falls below the traditional linewidth limit, their corresponding peaks (red and blue traces) merge (black trace) and become unresolvable.
    (b) Quantum circuits implementing conventional QVSA and WMDD protocols for detecting the frequency difference $\Delta \omega=\omega_2-\omega_1$.
    The WMDD protocol is implemented by applying a $\pi$ phase shift to the probe tone at time $T/2$.
    The QVSA mixer is realized by a motional Raman transition and heterodynes the quadrupolar probe and unknown signal tones, resulting in a displacement of the QHO \cite{Wu2025}.
    Expected (c) conventional and (e) WMDD spectroscopy generated by scanning the probe detuning ($\delta$) at several different $\Delta \omega$.
    (d, f) Expected mean phonon number $\langle n \rangle$ under the conventional protocol (d) and the WMDD protocol (f) as a function of $\Delta \omega$ at zero detuning $\delta = 0$. Shaded regions represent the uncertainty $\sigma_{\langle n \rangle}$ due to phonon shot noise for $N=1000$ trials.
    (d) Using the conventional protocol, the dominant source of shot noise in the measured $\langle n \rangle$ is independent of $\Delta\omega$, obscuring any dependence of $\langle n \rangle$ on $\Delta\omega$ and rendering $\Delta\omega$ unresolvable.
    (f) Using the WMDD protocol, sensitivity to $\Delta \omega$ is recovered by nullifying the $\Delta \omega$-independent shot noise at $\delta=0$.
    Here, the interrogation time is $T=1 \textrm{ms}$.}
    \label{fig:super}
\end{figure*}

It is highly desirable to extend super-resolution capabilities to other platforms and interactions with higher bandwidths and sensitivities, but without the strict requirement of correlated phase measurements. 
Quantum harmonic oscillators (QHOs) are a promising platform for electric fields and related quantities~\cite{Hempel2013,Burd2019,Gilmore2021,Deng2024}. 
Notably, the recent advent of quantum vector signal analysis (QVSA) \cite{Wu2025} via motional Raman transitions has significantly expanded the potential of QHO sensing by enabling wideband operation for fundamental science (e.g., dark matter detection~\cite{Bradley2003}), sensing (e.g. acceleration measurements~\cite{Campbell2017}, quantum radar~\cite{Barzanjeh2015,Assouly2023}), and communication (e.g. quantum network \cite{Xiang2017}).
Here, we describe and experimentally demonstrate a general broadband method for achieving frequency super-resolution applicable to any QHO system.
Specifically, we use a trapped $^{40}\textrm{Ca}^+$ ion to detect a minimum resolvable frequency difference of $\Delta\omega/2\pi = 5.0(1.6)$~Hz between two initially ``unknown" incoherent electric fields oscillating near 100~MHz with an interaction time of only $1 \textrm{ ms}$ – a resolution over $200\times$ beyond than the spectral linewidth limit.
The total measurement time required to resolve this $5 \textrm{ Hz}$ difference is reduced by more than five orders of magnitude compared to conventional techniques.
Our protocol implements a Walsh-modulated dynamic decoupling (WMDD) sequence \cite{Ball2015} using motional Raman transitions to leverage phase information, oft-neglected in conventional amplitude-sensitive spectroscopic techniques, and null the $\Delta \omega$-independent shot noise to realize nonzero $F_c$ even as $\Delta \omega \rightarrow 0$.
Our ability to achieve super-resolution can thus be understood as a more effective encoding of $\Delta \omega$ into the quantum state for measurement.
To the best of our knowledge, our protocol is the first on any platform to achieve broadband super-resolution of an oscillating electric field.
This breakthrough sheds new light on quantum metrology, demonstrating sensing in regimes previously considered unmeasurable.

\section{Theory}
Conventional frequency sensing techniques typically operate by mapping frequency information onto the amplitude of a measured signal — for example, light intensity in absorption spectroscopy~\cite{Demtroeder2003}, magnetic-field power spectra with NV centers~\cite{Rotem2019}, or phonon detection for electric-field sensing with trapped ions ~\cite{Wu2025}.
However, as the measurement outcomes associated with the two distinct frequencies begin to overlap and merge into a single peak, the ability to resolve the constituent tones is significantly reduced and ultimately lost, as illustrated in Fig.~\ref{fig:super}(a).

To understand the protocol, suppose that two unknown oscillating, uniform electric fields are present at frequencies $\omega_1$ and $\omega_2$.
A probe tone at $\omega_p$ is applied in an electric quadrupole configuration to generate a motional Raman displacement and sense the unknown tones.
The Hamiltonian of the ion motion under these fields is
\begin{align}
    \frac{\hat{H}}{\hbar} &= \omega_{o} \hat{a}^{\dagger} \hat{a} \notag \\
    &+ \Omega_{d} \left( \hat{a}+\hat{a}^\dagger \right) \left(\cos{ \left(  \omega_1 t \right)} + \cos{ \left(  \omega_2 t+\Delta\phi \right)}\right)\notag \\
    &+ \Omega_{p} \left( \hat{a} + \hat{a}^{\dagger} \right)^{2}
    \cos{ \left(  \omega_p t \right)},
    \label{eq:Hamil_base}
\end{align}
where $\omega_o$ is the natural frequency of the QHO, $\Delta \phi$ is the relative phase between the two unknown tones,
$\Omega_{d} =  eV_{d} \eta/\hbar$, $\Omega_{p} = eV_{p}\eta^{2}/\hbar$, $V_d$ is the voltage of the unknown tones, $V_p$ is the probe voltage, and $\eta$ is the effective Lamb-Dicke parameter~\cite{Wu2025}.
For simplicity, the amplitudes of the two unknown tones are assumed to be equal (see SI for the general case).

If the quadrupolar probe tone is applied at $\omega_p = \omega_{\mu} + \omega_o + \delta$, where $\omega_{\mu}=(\omega_1+\omega_2)/2$
for some fixed $\Delta \phi$, a conventional QVSA protocol based on motional Raman effect results, creating a coherent state $\ket{\alpha}$ with displacement (see SI):
\begin{align}                   
    \alpha\left(\Delta \omega,\Delta\phi,T\right) &= i \Omega_\textrm{eff} \left(\frac{1-e^{i\left(\delta - \Delta\omega/2\right)T}}{\delta - \Delta\omega/2} \right.\nonumber\\
    &\left.+ e^{-\imath\Delta\phi} \frac{1-e^{i \left(\delta+\Delta\omega/2\right)T}}{\delta+\Delta\omega/2}\right).
    \label{eqn:DispTwoTones}
\end{align}
Here, $\Delta\omega=\left(\omega_2-\omega_1\right)$ is the frequency difference between the unknown tones,  $\Omega_{\textrm{eff}} = \Omega_d\Omega_d \omega_o/((\omega_\mu \pm\Delta \omega/2))^2-\omega_o^2)$ is the effective Rabi frequency, and $T$ is the interrogation time.
Assuming that $\Delta\phi$ fluctuates uniformly in $\left[0, 2\pi\right]$ between measurements but remains approximately constant during $T$, the state of the system is: 
\begin{align}
    \hat{\rho} &= \frac{1}{2\pi} \int{\ket{\alpha\left(\Delta \omega,\Delta\phi, T\right)}\bra{\alpha\left(\Delta \omega,\Delta\phi, T\right)}d \left(\Delta\phi\right)}.
    \label{eqn:density_system}
\end{align}
The value of $\Delta\omega$ is encoded into the complex displacement $\alpha$ and can be retrieved by measuring the mean phonon number $\langle n\rangle = \textrm{Tr}{\left( \hat{\rho}\hat{a}^+\hat{a} \right)} = \left( \left| \alpha \left(\Delta \omega,0, T\right) \right|^2 + \left|\alpha\left(\Delta \omega,\pi, T\right)\right|^2\right)/2$.
By scanning the probe detuning $\delta$, the spectrum of the unknown tones can be characterized at different $\Delta\omega$ (Fig.~\ref{fig:super}(c)). 

However, in the super-resolution limit where $\Delta\omega T \ll 1$, 
it is straightforward to show that $\left|\alpha\left(\Delta \omega,0,T\right)\right|^2 \approx 4\Omega_{\textrm{eff}}^2 T^2 - \frac{1}{3}\Omega_{\textrm{eff}}^2 T^4\Delta\omega^2$ and $\left|\alpha\left(\Delta \omega,\pi, T\right)\right|^2 \approx \frac{1}{4}\Omega_{\textrm{eff}}^2 T^4\Delta\omega^2$.
In the conventional protocol, the uncertainty of $\Delta\omega$ under the  shot noise limit is thus $\sigma_{\Delta\omega} = \frac{\sigma_n}{\partial n/\partial \Delta\omega} = 
\frac{12\sqrt{2}}{\Omega_{\textrm{eff}} T^3 \Delta\omega}$, which diverges as $\Delta\omega\rightarrow0$, recovering the traditional linewidth limit where frequencies become unresolvable.
This divergence results from the shot noise introduced by the $\Delta\omega$-independent displacement, $2\Omega_{\textrm{eff}} T$, which dominates and obscures any dependence on $\Delta \omega$, shown in Fig.~\ref{fig:super}(d).
To achieve frequency super-resolution, this $\Delta\omega$-independent displacement must therefore be removed.

\begin{figure*}
    \centering
    \includegraphics[width =1\textwidth]{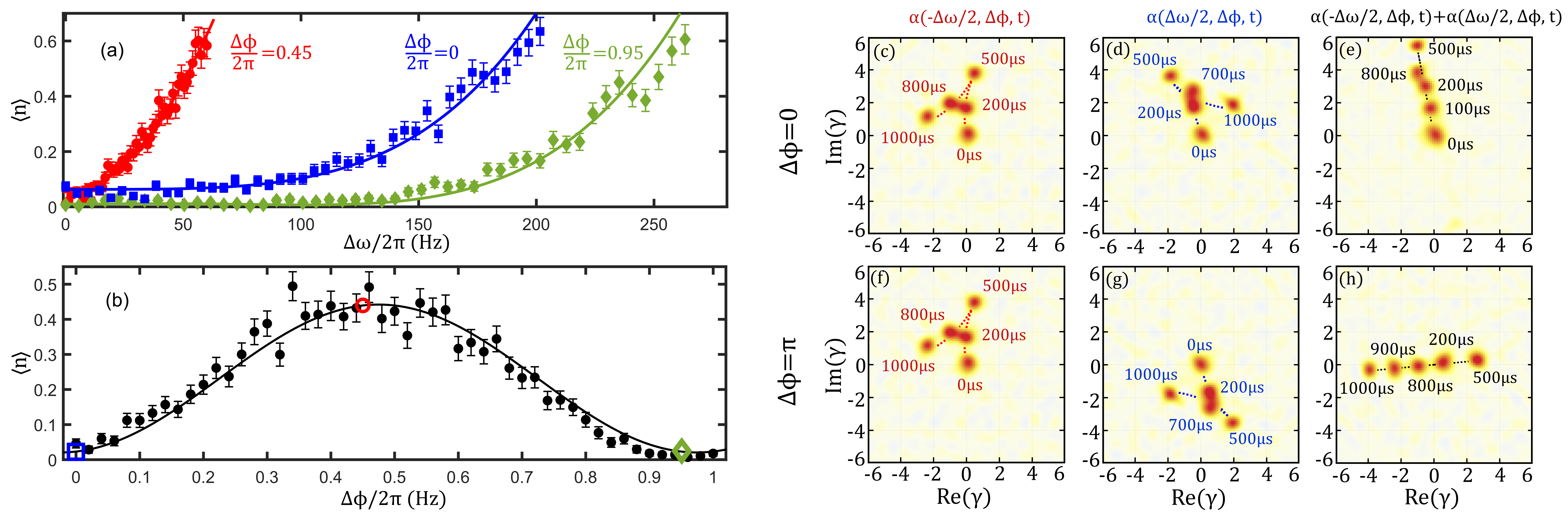}
    \caption{
    Dependence of the WMDD protocol on the relative phase $\Delta \phi$ between two signal tones $\omega_1 \approx \omega_2 \approx 2\pi \cdot 86 \textrm{ MHz}$.
    (a) Super-resolution spectrum of $\Delta\omega = \omega_2 - \omega_1$ for $\Delta \phi=0.9\pi$ (red, circle), $\Delta \phi=1.9\pi$ (green, diamond), and $\Delta \phi=0$ (blue, square).
    (b) Measured $\langle n \rangle$ as a function of $\Delta \phi$ at $\Delta\omega/2\pi=50 \textrm{ Hz}$.
    Experimental data (dotted points) are in excellent agreement with fits to Eq.~\eqref{eq:phase} (solid lines).
    Almost all parameters used in the fit are independently measured or controlled (e.g. $\Omega_{\textrm{eff}}$ and $T$), with a uniform vertical offset being the only fitting parameter.
    No additional vertical or horizontal scaling is applied.
    (a-b) A total interrogation time $T = 1\textrm{ ms}$ was used.
    Error bars represent one standard error. Each point is averaged over $N=600$ repetitions.
    (c-h) Phase space evolution of the QHO Wigner function under the WMDD protocol for $\Delta\phi=0$ (c, d, e) and $\Delta\phi=\pi$ (f, g, h) at $\Delta \omega/2\pi=200 \textrm{ Hz}$.
    The Wigner function is experimentally measured using motional state tomography (Ref.~\cite{Fluhmann2020}).
    The blue (red) dashed line represent the analytical phase-space trajectory due to evolution under a single tone. 
    The black dashed lines show the resultant trajectory when both tones are applied. 
    (e,h) Following phase reversal at $T=500 \mu s$, the motional trajectory retraces itself to remove the $\Delta\omega$-independent displacement, leaving only a $\Delta\omega$-dependent displacement.
    }
    \label{fig:scan}
\end{figure*}
To effectively cancel this $\Delta\omega$-independent displacement, we apply Walsh-Modulated Dynamical Decoupling (WMDD).
The total displacement under WMDD is:
\begin{align}    
    \label{eq:alpha_WMDD}
    \alpha_{\textrm{SR}} \left(\Delta\omega, \Delta\phi, T\right)&=\alpha\left(\Delta\omega, \Delta\phi, T/2\right) \\ \nonumber
    & +e^{i\pi} \left(\alpha\left(\Delta\omega, \Delta\phi, T/2\right)-\alpha\left(\Delta\omega, \Delta\phi, T\right)\right)
\end{align}
The resultant mean phonon number $\langle n \rangle$ for $\delta=0$ is (see SI):
\begin{align}
    \langle n \rangle = \frac{256\Omega_\textrm{eff}^2}{\Delta\omega^2} \sin^4{\left(\frac{\Delta\omega T}{8}\right)} \sin^2{\left(\frac{\Delta\omega T/2-\Delta\phi}{2}\right)}.
    \label{eq:phase}
\end{align}
The cancellation of excess shot noise can be understood by examining the evolution of the Wigner function in phase space \cite{Fluhmann2020}.
Assuming the QHO is initialized in the vacuum state $\ket{0}$, when $\Delta\omega = 0$, the Wigner function of the QHO is translated along a fixed axis in phase space (Fig.~\ref{fig:scan}(e, h)).
If the direction of this displacement is reversed, by e.g. shifting the probe phase by $\pi$ at $T/2$, the motional trajectory retraces itself and returns to the origin, thereby canceling the $\Delta\omega$-independent displacement, shown in Fig.~\ref{fig:super}(e) for $\delta=0$.
For nonzero $\Delta\omega$, this cancellation is incomplete due to accumulation of a phase $\Delta\omega T/2$ prior to reversal, leaving a residual displacement dependent solely on $\Delta\omega$, illustrated in Fig.~\ref{fig:scan}(c–h).
As a result, $\Delta\omega$ becomes `encoded' into the mean phonon number $\langle n \rangle$, which can be efficiently measured in a QHO system, shown in Fig.\ref{fig:super}(f). 
In the super-resolution limit $\Delta\omega T\ll 1$
with random phase $\Delta \phi$, Eq.~\eqref{eq:phase} becomes
\begin{align}
    \langle n \rangle_{\textrm{SR}} = \langle \left|\alpha_{\textrm{SR}}\left(\Delta\omega\right) \right|^2 \rangle_{\Delta\phi} \approx  \frac{\Omega_\textrm{eff}^2}{32} \Delta\omega^2 T^4.
    \label{eq:incoherent}
\end{align}
The corresponding upper bound of the quantum Fisher information is
$\mathcal{F}(\Delta\omega)
\leq (\Omega_{\textrm{eff}}T^2)^2/8$ (see SI).
Notably, this result is independent of $\Delta\omega$, resulting in an uncertainty $\sigma_{\Delta\omega}$ that does not diverge as $\Delta\omega \rightarrow 0$, thereby enabling super-resolution measurement. 
Conceptually, WMDD operates in phase space in a manner analogous to zero-delay Ramsey spectroscopy on the Bloch sphere. 
$\mathcal{F}_Q(\Delta \omega)$ effectively takes a form similar to that of Ramsey spectroscopy \cite{Wolf2019}, suggesting a potential for ultra-performant frequency sensing with our approach.

\section{Discussion}
To verify and demonstrate our technique, we utilize a QHO realized on the axial motional mode of a single trapped \ca\ ion with secular frequency $\omega_o \approx 2\pi \cdot 700 \textrm{ kHz}$.
The ground motional state of the axial mode is initially prepared via Doppler cooling of the ion followed by resolved-sideband cooling on the optical $^{2}S_{1/2} \leftrightarrow ^{2}D_{5/2}$ qubit transition to yield a mean phonon number $\langle n \rangle \approx 0.01(1)$ (see Ref.~\cite{Wu2025,WuH2025} for more detail).

For all experiments, the two `unknown' signal tones are applied as uniform electric fields by coupling voltages to the trap electrodes in a dipole configuration with frequencies $\omega_i = \omega_\mu \mp \Delta\omega/2$ for $i = 1,2$ and relative phase $\Delta\phi$.
Prior to application of the super-resolution protocol, $\omega_{\mu}$ is first obtained using conventional QVSA spectroscopy (see SI).
During super-resolution protocol, the probe frequency is set to $\omega_p = \omega_{\mu}+\omega_o$, with a phase shift of $\pi$ applied to the probe tone at $T/2$.
Following detection, the residual displacement is determined using either an inversion estimator for fixed $\Delta\phi$, or a phase-averaged estimator when $\Delta\phi$ varies randomly between measurements (see SI).

As a first test, the super-resolution protocol is applied for $\Delta \phi$ fixed at several values (0, 0.9$\pi$, and $1.9\pi$).
The measured $\braket{n}$ for $\omega_\mu = 2\pi \cdot86 \textrm{ MHz}$ and $T = 1 \textrm{ ms}$ is shown in Fig.~\ref{fig:scan}~(a) as a function of $\Delta\omega$, where lines depict the analytical result given by Eq.~\eqref{eq:phase}.
All parameters in Eq.~\eqref{eq:phase} are either calibrated experimentally (see SI) or set accurately in hardware.
To account for instability of the QHO frequency between measurements, we add a uniform offset when fitting Eq.~\eqref{eq:phase} as the only free parameter.
Fig.~\ref{fig:scan}(b) further shows the measured $\braket{n}$ as a function of $\Delta\phi$ at fixed $\Delta\omega = 2\pi \cdot 50  \textrm{ Hz}$.
The excellent agreement of experimental results with the theory indicate that, with only straightforward analysis, $\Delta\omega$ can be measured orders of magnitude below the traditional linewidth limit ($\sim 1/T=1 \textrm{ kHz}$) for fixed $\Delta\phi$.

\begin{figure}[h]
    \centering
    \includegraphics[width = 0.5\textwidth]{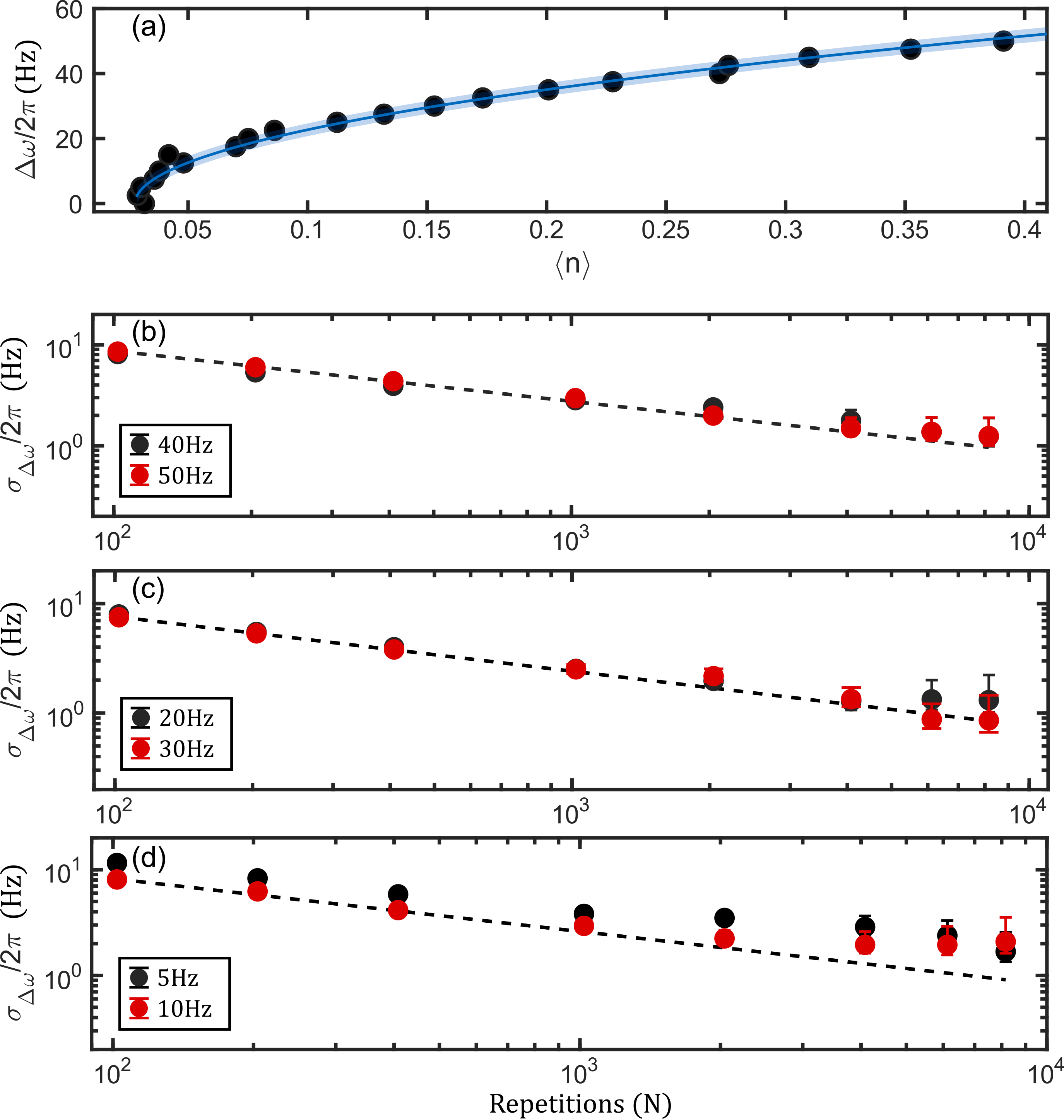}
    \caption{
    Characterizing ultimate resolution of $\Delta\omega$ with two incoherent signal tones centered at $2\pi \cdot86  \textrm{ MHz}$.
    (a) Calibration curve used to convert the measured $\langle n\rangle$ to the parameter $\Delta \omega$.
    Experimental data is plotted as dotted points and fitted to an inversion of the analytical model (solid lines) in Eq.~\eqref{eq:incoherent} with only a constant horizontal offset as the sole fit parameter.
    The colored band indicates the $68\%$ prediction band.
    Each point is averaged over $N=5000$ measurements.
    There is no additional rescaling.
    (b-d) Overlapping Allan deviations of $\Delta \omega$ at different frequency separations $\Delta \omega$ with fixed known Rabi frequency $\Omega_{\textrm{eff}}$. 
    (b) $\Delta\omega/2\pi = 40,\; 50 \textrm{ Hz}$.
    (c) $\Delta\omega/2\pi = 20,\; 30 \textrm{ Hz}$.
    (d) $\Delta\omega/2\pi = 5,\; 10 \textrm{ Hz}$.
    (a-d) Data was taken using a total interrogation time $T=1 \textrm{ ms}$. To simulate incoherence, the relative phase between the two signal tones $\Delta \phi$ was varied randomly between each measurement from a uniform distribution on $\left[0, 2\pi \right]$ .
    Error bars represent one standard error. 
    The dashed lines represent a $1/\sqrt{N}$ trend.
    Each repetition takes $\approx 13~\textrm{ms}$.
    }
    \label{fig:allan}
\end{figure}

Next, to test a more general scenario where the two signal tones are incoherent, $\Delta\phi$ is varied randomly between each measurement over a uniform distribution on $\left[0, 2\pi \right]$. 
Using the phase-averaged estimator, the relationship between $\Delta\omega$ and the mean phonon number $\langle n \rangle$ for two incoherent signal tones is characterized, shown in Fig.~\ref{fig:allan}(a).
Here, $\Omega_{\textrm{eff}}$ is independently measured and used in the fit.
The solid line represents a fit of the data to inversion function of Eq.~\eqref{eq:incoherent} with a uniform horizontal offset as the sole free parameter.
The shaded regions present the $68\%$ prediction band for $\Delta \omega$ from the data.
These narrow prediction bands, as in Fig.~\ref{fig:super}(f) evidence clearly the ability of the protocol to achieve super-resolution on even incoherent signals.

Following this demonstration, an Allan deviation is performed at six different values of $\Delta\omega$ to evaluate the ultimate achievable resolution for $\Delta\omega$ with known $\Omega_{\textrm{eff}}$ from pre-calibration, presented in Fig.~\ref{fig:allan}(b-d).
Here, the dashed lines indicate the expected $1/\sqrt{N}$ scaling.
Deviations from this trend result from systematic sources of error, and are chiefly due to instability in the natural frequency of the QHO.
After $N = 8160$ repetitions, the uncertainty in $\Delta\omega$ falls below the single-Hz level, and is nearly identical across all values of $\Delta \omega$, consistent with the theoretical prediction that, in the super-resolved measurement regime, the uncertainty is independent of $\Delta\omega$.
To resolve signals separated by $5 \textrm{ Hz}$, our WMDD protocol requires a measurement time of $110 \textrm{ s}$.
By comparison, spectroscopy using a conventional protocol would require roughly a measurement time of $3\cdot10^7 \textrm{ s}$ to achieve a similar resolution - this is a result that would be practically inachievable at any level without even factoring in systematic sources of error such as long-term drifts.

\begin{figure}[h]
    \centering
    \includegraphics[width = 0.5\textwidth]{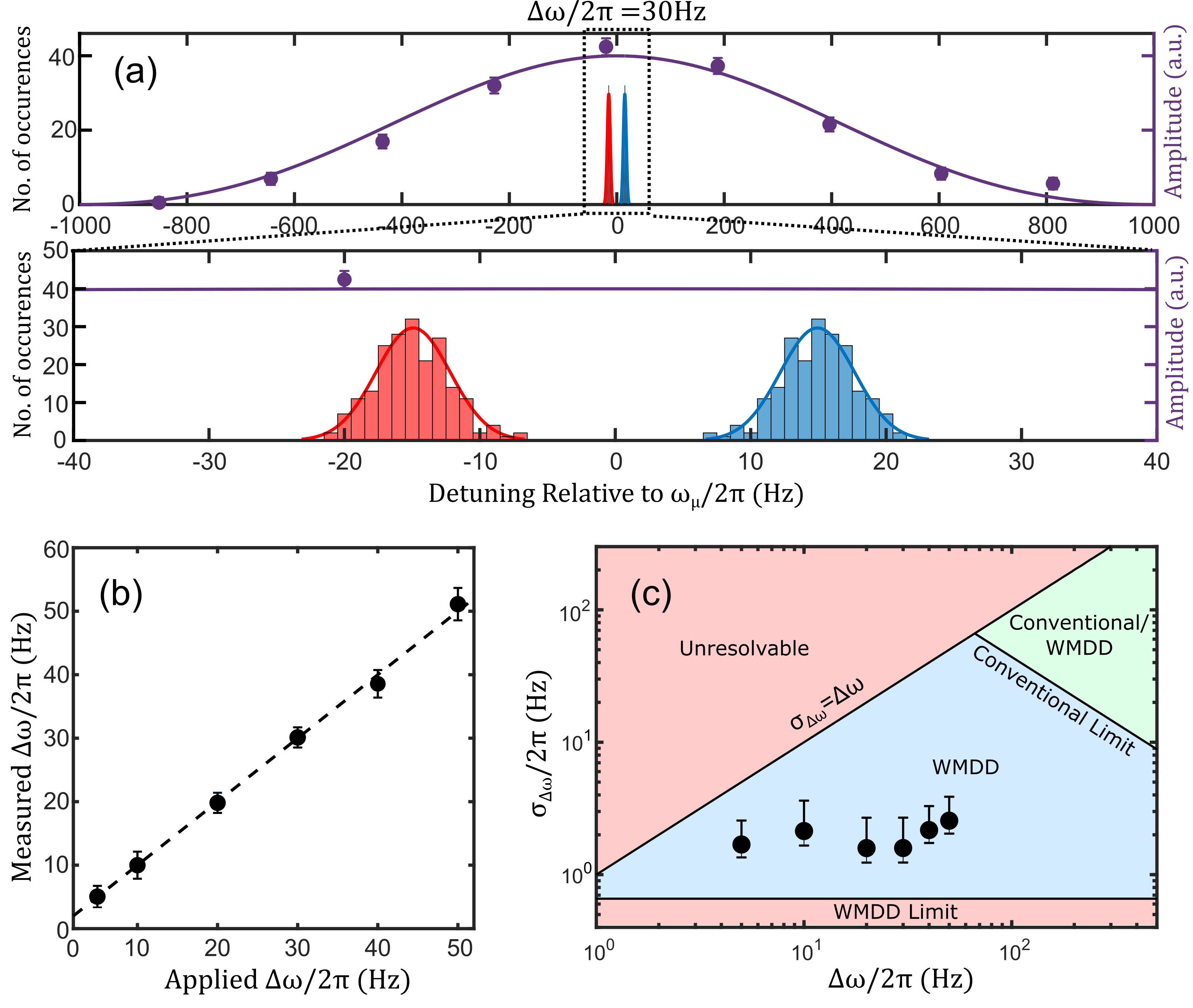}
    \caption{Measurement of ``unknown" $\Delta\omega$ with two incoherent signal tones at $\omega_{\mu} = 2\pi\cdot 86 \textrm{ MHz}$ with all sources of error accounted for.
    (a) Distribution of measurements of $\omega_{\mu}\pm \Delta\omega/2$ for $\Delta\omega/2\pi=30 \textrm{ Hz}$.
    For comparison, measurements (purple circles) and a corresponding fit (purple line) are shown for a conventional protocol.
    The solid red and blue lines depict Poisson fits to the measured distributions.
    Each occurrence is averaged over $N=102$ repetitions.
    (b) Measured values of $\Delta\omega$ compared against the true values.
    The dashed line represents an ideal 1:1 correspondence between the applied and measured values.
    (c) Comparison of resolvable regions between conventional and WMDD protocols.
    The region resolvable by both conventional and WMDD protocols is bounded by the minimum uncertainty achievable with conventional protocols - conventional limit (green).
    The WMDD protocol extends this region and is bounded by the QFI - the WMDD limit (blue).
    Beyond $\sigma_{\Delta\omega}=\Delta\omega$, $\Delta\omega$ becomes unresolvable (red).
    At $\Delta\omega/2\pi = 5 \textrm{ Hz}$, the WMDD protocol requires $10^{5}$ fewer repetitions than a conventional protocol to achieve the same uncertainty.
    (b-c) An interrogation time of $T = 1\textrm{ ms}$ is used with $N=8160$ repetitions. Error bars represent one standard error.
    }
    \label{fig:X2}
\end{figure}

As a final demonstration to characterize the precision and accuracy of our super-resolution protocol, we apply and sense two unknown, incoherent signals.
We first optimize the probe frequency $\omega_{p}$ by scanning the probe detuning $\delta$ with our WMDD protocol and minimizing the measured mean phonon number $\langle n \rangle$.
The effective Rabi frequency $\Omega_{\textrm{eff}}$ is extracted by performing conventional spectroscopy, 
with a resulting prediction error ($\sigma_{\Omega_{\textrm{eff}}}$) $\approx 4\%$ based on measured phonon (see SI).
Next, we conduct a super-resolution measurement with interrogation time $T = 1 \textrm{ ms}$ at the optimized $\omega_p$. to determine $\Delta\omega$.
To suppress the effect of drifts in $\omega_o$, a reference signal with known $\Delta\omega$ is concurrently measured. 
The distribution of the measured $\omega_\mu \pm\Delta\omega/2$ is shown in Fig.~\ref{fig:X2}(a) for $\Delta\omega = 2\pi \cdot30 \textrm{ Hz}$ (see SI for further data at different $\Delta\omega$).
As expected, two incoherent spectral components (blue and red) are clearly resolved at frequencies $\omega_{\mu}\pm \Delta\omega/2$, in contrast to the conventional method, shown in purple.
This process is repeated at several values of $\Delta\omega$ to compare the measured values $\Delta\omega$ to their applied true values in Fig.~\ref{fig:X2}(b).
The excellent direct correspondence of our measurements to the true values in Fig.~\ref{fig:X2}(b) demonstrates the accuracy and reliability of our super-resolution protocol.
To contextualize the frequency resolution achievable with our super-resolution protocol, we compare our resolution limits against those of conventional protocol, depicted in Fig.~\ref{fig:X2}(c).
The $\sigma_{\Delta\omega}=\Delta\omega$ line demarcates the resolvable from the unresolvable regions.
The uncertainty using conventional protocol crosses this bound at $\Delta\omega \approx 2\pi \cdot66  \textrm{ Hz}$ using an interrogation time $T = 1 \textrm{ ms}$ and $N = 8160$ repetitions.
In contrast, the super-resolution limit $\sigma_{\Delta\omega}^{\left(\textrm{SR}\right)} = 2\pi \cdot0.65  \textrm{ Hz}$ is independent of $\Delta\omega$ and reaches the $\sigma_{\Delta\omega}$ limit at  $\Delta\omega \leq 2\pi\cdot 1 \textrm{ Hz}$ for the same parameters.
Our super-resolution protocol has thus demonstrably extended the spectrally resolvable frequency range by a factor of $\approx 14\times$; in principle, this range could be extended in total by over two orders of magnitude.
Deviations of the measured uncertainty from $\sigma_{\Delta\omega}^{\left(\textrm{SR}\right)}$ arise from a combination of unsaturated estimators \cite{Wu2025}, uncertainty in $\Omega_{\textrm{eff}}$ and $\omega_\mu$, and systematic errors due to drifts in $\omega_o$ imperfections when applying the WMDD protocol.
Notably, our protocol achieves at $\Delta\omega/2\pi = 5 \textrm{ Hz}$ a resolution that requires $>10^5$ fewer experimental repetitions than the conventional approach, without considering even systematic sources of error in the latter.
This enhanced resolution demonstrated by our protocol allows determination of $\Delta \omega$ at levels 200 times beyond the spectral linewidth limit ($\sim 1/T = 1 \textrm{ kHz}$).

\section{Summary and outlook}
In summary, we have shown that it is possible to resolve two oscillating electric-fields at a level $200 \times$ beyond the spectral linewidth limit, a regime traditionally considered inaccessible.
This technique relies on a heterodyned quantum interaction that leverages oft-neglected phase information to actively suppress excess shot noise via a dynamical decoupling sequence.
Since this method is based on motional Raman interactions, it retains sensitivity across a wide frequency range.
This approach offers a broadly applicable solution for achieving frequency super-resolution across both the radio-frequency and the microwave domains \cite{Wu2025}.
Several straightforward improvements would yield further gains in resolution.
By increasing both the interrogation time and the input power of the probe tone, the detection range can be extended by at least an additional order of magnitude.
Our super-resolution protocol is further compatible with quantum amplification via non-classical states \cite{Wolf2019,McCormick2019,Burd2019}, offering enhancements to the precision of our super-resolution protocol, particularly on short timescales.

Finally, the impacts of our protocol are wide-ranging.
Since the protocol is equivalent to motional Ramsey spectroscopy, it can be readily integrated into existing low-depth variational quantum circuits to enable further optimization of input states, allowing for more efficient and programmable quantum sensing~\cite{Kaubruegger2021,Marciniak2022}, in even noisy environments~\cite{Meyer2021}.
Further, the protocol is straightforwardly adaptable to a variety of QHO platforms, such as cavity QED~\cite{Schlawin2022}, and superconducting circuits~\cite{Krantz2019}. 
Our protocol can be implemented using other Raman-based techniques; for instance, with NV centers for broadband magnetic field sensing as demonstrated in Ref.~\cite{Wang2022}, or coupling with alternative two-level systems aimed at achieving frequency super-resolution.
Finally, our super-resolution capability opens the possibility of increasing the number of communication channels and improving spectral efficiency in bandwidth-limited communication environments, such as in next-generation wireless networks, satellite links, and quantum communication systems~\cite{Xiang2017}.

\section{Acknowledgments}
This work was supported by the NSF (PHY-2110421 and OMA-2016245), AFOSR (130427-5114546), and the ARO (W911NF-19-1-0297).

\section{Competing interests}
The authors declare no competing interests.

\bibliography{ref}
\appendix
\onecolumngrid
\include{SI}

\end{document}

%% file: SI.tex


\section{Supplementary Information}

\section{Appendix A: Displacement due to a motional Raman transition }
Suppose an unknown, oscillatory signal at frequency $\omega_d$ is applied to the trap in a dipole configuration while a signal at frequency at $\omega_q = \omega_d + \delta + \omega$ are applied to the trap in a quadrupole configuration.
We assume $\omega_d \gg \omega\gg \delta$.
The motion of the ion can described by the Hamiltonian
\begin{align}
    \hat{H}/\hbar =&  \omega\hat{a}^\dagger\hat{a} + \frac{eV_q}{2 \hbar r_o^2}\frac{\hbar}{2m\omega} (\hat{a}+\hat{a}^\dagger)^2
    \cos{\left( \left(\omega_d + \delta + \omega \right) t \right)} + \frac{\kappa eV_d}{\hbar  r_o}\sqrt{\frac{\hbar}{2m\omega}}(\hat{a}+\hat{a}^\dagger) \cos{ \left( \omega_d t+\phi_d \right) }\nonumber\\
    =&  \omega\hat{a}^\dagger\hat{a} + \Omega_q (\hat{a}+\hat{a}^\dagger)^2
    \cos{\left( \left(\omega_d + \delta + \omega \right) t \right)} + \Omega_d(\hat{a}+\hat{a}^\dagger) \cos{ \left( \omega_d t+\phi_d \right) },
\end{align}
where $\Omega_q = eV_q x_o^2/(\hbar r_o^2)$, $\Omega_d = \kappa eV_d x_o/(\hbar  r_o)$, $r_o$ is half the distance between the trap's radial electrodes, $x_o=\sqrt{\hbar/(2m\omega)}$ is the zero-point wavefunction of the QHO, and $\kappa$ is a geometrical factor. 

In the interaction picture with respect to the harmonic oscillator we have:
\begin{align}
    \hat{H}/\hbar =& \Omega_q \left( \hat{a}^2 e^{-2\imath\omega t}+2\left(\hat{a}^\dagger\hat{a} + \frac{1}{2}\right)+ \hat{a}^{\dagger 2} e^{2\imath\omega t} \right) \cos{\left( \left(\omega_d + \delta + \omega \right) t \right)} \nonumber \\
    &+ \Omega_d \left( \hat{a} e^{-\imath\omega t} +\hat{a}^\dagger e^{\imath \omega t} \right) \cos{\left( \omega_d t+\phi_d \right)}
\end{align}

The Magnus Expansion gives the propagator as $\hat{U} = e^{\sum_i \hat{A}_i}$.
The first order Magnus term is:
\begin{align}
    \hat{A}_1 = -\frac{\imath}{\hbar}\int_{t_o}^t \hat{H}(t') dt' 
\end{align}
and leads to light shifts of the ion motional states.
These light shifts oscillate at $\sim\omega_d$ with amplitudes that scale as $\Omega_q/\omega_d$ and $\Omega_d/\omega_d$.
We assume that these amplitudes are small and ignore this term -- in practice, any effect of the light shift can be accounted for by a slight change in $\omega$.

The second order Magnus term is given as: 
\begin{align} \label{eq:magnus_2}
    \hat{A}_2 = \frac{1}{2}&\left(\frac{-\imath}{\hbar}\right)^2 \int_{t_o}^t\int_{t_o}^{t'} \left[ \hat{H}(t'),\hat{H}(t'')\right] dt' dt''.
\end{align}

This order will also contain terms that give rise to a light shift that arise from the commutator of the $\Omega_ q$ terms at $t'$ with $\Omega_q$ terms at $t''$ as well as from the commutator of the $\Omega_d$ terms at $t'$ with $\Omega_e$ terms at $t''$.
These terms will oscillate at $\omega_d$ with amplitudes that scale as $\Omega_q^2/\omega_d^2$ and $\Omega_d^2/\omega_d^2$, respectively.
They lead to the intrinsic squeezing in QVSA.
As with the first order light shift terms, we will ignore them using what is essentially a rotating wave approximation. 
The remaining terms of $\hat{A}_2$ are due to the commutators between the $\Omega_q$ and $\Omega_d$ terms.
These result in the most important evolution of the system and are given as:
\begin{align}
    &\left[ \hat{H}(t'),\hat{H}(t'')\right]  \approx  \Omega_q\Omega_d 
    \cos{\left( \left(\omega_d + \delta + \omega \right) t' \right)}
    \cos{\left( \omega_d t''+ \phi \right)}
    \left(  e^{-2\imath\omega t'} e^{\imath\omega t''} \left[ \hat{a}^2,\hat{a}^\dagger \right] \right.  +2 e^{-\imath\omega t''}\left[\hat{a}^\dagger\hat{a},\hat{a}\right] \nonumber \\
    &+ 2 e^{\imath\omega t''}\left[\hat{a}^\dagger\hat{a},\hat{a}^\dagger\right] \left.+ e^{2\imath\omega t'} e^{-\imath\omega t''}\left[\hat{a}^{\dagger 2},\hat{a}\right]\right) 
    +
    \Omega_d\Omega_q 
    \cos{\left( \omega_d t' + \phi\right)}
    \cos{\left( \left(\omega_d + \delta + \omega \right) t'' \right)}
    \left(  e^{-2\imath\omega t''} e^{\imath\omega t'}\left[\hat{a}^\dagger,\hat{a}^2 \right] \right. \nonumber\\
    &+2 e^{-\imath\omega t'}\left[\hat{a},\hat{a}^\dagger\hat{a}\right]  + 2 e^{\imath\omega t'}\left[\hat{a}^\dagger, \hat{a}^\dagger\hat{a}\right] 
    \left.+ e^{2\imath\omega t''} e^{-\imath\omega t'}\left[\hat{a},\hat{a}^{\dagger 2}\right]\right).
\end{align}
Now, using the relations:
\begin{align}
    \left[\hat{a}^2,\hat{a}^\dagger\right] &= 2\hat{a}\nonumber\\
    \left[\hat{a}^\dagger\hat{a},\hat{a}\right] &= -\hat{a}\nonumber\\
    \left[\hat{a}^\dagger\hat{a},\hat{a}^\dagger\right] &= \hat{a}^\dagger\nonumber\\
    \left[\hat{a}^{\dagger 2},\hat{a}\right] &= -2\hat{a}^\dagger
\end{align}

we have: 
\begin{align} 
    &\left[ \hat{H}(t'),\hat{H}(t'')\right]  \approx  2\Omega_q\Omega_d 
    \cos{\left( \left(\omega_d + \delta + \omega \right) t' \right)}
    \cos{\left( \omega_d t'' + \phi\right)}
    \left(  e^{-2\imath\omega t'} e^{\imath\omega t''}\hat{a}\right. - e^{-\imath\omega t''}\hat{a} +  e^{\imath\omega t''}\hat{a}^\dagger \left.- e^{2\imath\omega t'} e^{-\imath\omega t''}\hat{a}^\dagger\right)\nonumber\\
    &- 
    2\Omega_q\Omega_d 
    \cos{\left( \omega_d t' + \phi\right)}
    \cos{\left( \left(\omega_d + \delta + \omega \right) t'' \right)}
    \left(  e^{-2\imath\omega t''} e^{\imath\omega t'}\hat{a} - e^{-\imath\omega t'}\hat{a} +  e^{\imath\omega t'}\hat{a}^\dagger - e^{2\imath\omega t''} e^{-\imath\omega t'}\hat{a}^\dagger\right)\nonumber\\
    & \approx 2\Omega_q\Omega_d 
    \cos{\left( \left(\omega_d + \delta + \omega \right) t' \right)}
    {\left( \omega_d t'' + \phi\right)}
    \left(  \hat{a}\left(e^{-2\imath\omega t'} e^{\imath\omega t''}- e^{-\imath\omega t''}\right) +  \hat{a}^\dagger\left(e^{\imath\omega t''} - e^{2\imath\omega t'} e^{-\imath\omega t''}\right)\right)\nonumber\\
    &- 2\Omega_q\Omega_d 
    \cos{\left( \omega_d t' + \phi\right)}
    \cos{\left( \left(\omega_d + \delta + \omega \right) t'' \right)} \left(  \hat{a}\left(e^{-2\imath\omega t''} e^{\imath\omega t'}- e^{-\imath\omega t'}\right) +  \hat{a}^\dagger\left(e^{\imath\omega t'} - e^{2\imath\omega t''} e^{-\imath\omega t'}\right)\right)
\end{align}

To simplify this, we need to make some rotating wave approximations. 
First, let's examine the cosines:
\begin{align}
    \cos{\left( \left(\omega_d + \delta + \omega \right) t' \right)} \cos{\left( \omega_d t'' + \phi \right)} &= \frac{1}{4}\left(e^{\imath(\omega_d + \delta + \omega ) t'} + e^{-\imath(\omega_d + \delta + \omega ) t'}\right)\left( e^{\imath(\omega_d t''+ \phi )}+ e^{-\imath(\omega_d t'' + \phi )}\right)\nonumber\\
    &\approx \frac{1}{4}\left(e^{\imath( \omega_d(t'-t'') + (\delta + \omega)t' -\phi)} + e^{-\imath( \omega_d(t'-t'') + (\delta + \omega)t' -\phi)}\right)\nonumber\\
    &\approx \frac{1}{2}\cos{(\omega_d(t'-t'') + (\delta + \omega)t' -\phi)},
\end{align}
where we've dropped terms that will end up oscillating at $2\omega_d$.
Similarly,
\begin{align}
    \cos{\left( \left(\omega_d + \delta + \omega \right) t'' \right)} \cos{\left( \omega_d t' + \phi \right)} \approx \frac{1}{2}\cos{(\omega_d(t''-t') + (\delta + \omega)t'' -\phi)},
\end{align}

Now, we evaluate the four terms.
First:
\begin{align}
    &\Omega_q\Omega_d \hat{a} \cos{(\omega_d(t'-t'') + (\delta + \omega)t' -\phi)} \left(e^{-2\imath\omega t'} e^{\imath\omega t''}- e^{-\imath\omega t''}\right) \nonumber\\
    &\approx \hat{a}\frac{\Omega_q\Omega_d}{2}\left(e^{\imath(\omega_d(t'-t'') + \delta t' + \omega(t''-t'))-\phi)} - e^{\imath(\omega_d(t'-t'') + \delta t' - \omega(t''-t'))-\phi)} \right)\nonumber\\
    &\approx \hat{a}\frac{\Omega_q\Omega_d}{2}e^{\imath(\omega_d(t'-t'') + \delta t' -\phi)}\left(e^{\imath\omega(t''-t')} - e^{-\imath\omega(t''-t')} \right)\nonumber\\
    &\approx \imath\Omega_q \Omega_d \hat{a} e^{\imath(\omega_d (t'-t'') + \delta t' -\phi)} \sin({\omega(t''-t')}),
\end{align}
where we just take the $e^{+\imath}$ term from the cosine and neglect the terms that would effectively oscillate at $2\omega$.

Second:
\begin{align}
    &\Omega_q\Omega_d \hat{a}^\dagger \cos{(\omega_d(t'-t'') + (\delta + \omega)t' -\phi)} \left(e^{\imath\omega t''} - e^{2\imath\omega t'} e^{-\imath\omega t''}\right) \nonumber\\
    &\approx \hat{a}^\dagger\frac{\Omega_q\Omega_d}{2}\left(e^{-\imath(\omega_d(t'-t'') + \delta t' + \omega(t'-t'')-\phi)} - e^{-\imath(\omega_d(t'-t'') + \delta t' - \omega(t'-t'')-\phi)}\right)\nonumber\\ 
    &\approx \hat{a}^\dagger\frac{\Omega_q\Omega_d}{2} e^{-\imath(\omega_d(t'-t'') + \delta t' -\phi)} \left(e^{-\imath \omega(t'-t'')} - e^{\imath(\omega(t'-t'')}\right)\nonumber\\ 
    &\approx \imath\Omega_q \Omega_d \hat{a}^\dagger e^{-\imath(\omega_d (t'-t'') + \delta t' -\phi)} \sin({\omega(t''-t')}),
\end{align}
where we just take the $e^{-\imath}$ term from the cosine and neglect the terms that would effectively oscillate at $2\omega$.

The third and fourth are the same as the first and second term with$t'$ and $t''$ interchanged, respectively.
Thus, we have:
Third:
\begin{align}
    &\Omega_q\Omega_d \cos{(\omega_d(t''-t') + (\delta + \omega)t'' -\phi)}
    \hat{a}\left(e^{-2\imath\omega t''} e^{\imath\omega t'}- e^{-\imath\omega t'}\right) \nonumber\\
    &\approx \imath\Omega_q \Omega_d \hat{a} e^{\imath(\omega_d (t''-t') + \delta t'' -\phi)} \sin({\omega(t'-t'')}).
\end{align}
Fourth:
\begin{align}
    &\Omega_q\Omega_d \cos{(\omega_d(t''-t') + (\delta + \omega)t'' -\phi)}\hat{a}^\dagger\left(e^{\imath\omega t'} - e^{2\imath\omega t''} e^{-\imath\omega t'}\right) \nonumber\\
    &\approx \imath\Omega_q \Omega_d \hat{a}^\dagger e^{-\imath(\omega_d (t''-t') + \delta t'' -\phi)} \sin({\omega(t'-t'')}).
\end{align}

Putting it all together we have:
\begin{align}
    \hat{a}\int_{t_o}^t\int_{t_o}^{t'} \sin{(\omega(t''-t'))}\left(e^{\imath(\omega_d (t'-t'') + \delta t' -\phi)}
    + e^{\imath(\omega_d (t''-t') + \delta t'' -\phi)}\right)dt'' dt' &\approx \imath \hat{a} e^{-\imath\phi} \frac{(e^{\imath\delta t_o}-e^{\imath \delta t})\omega(\delta^2 - 2\omega^2 + 2\delta\omega_d + 2\omega_d^2)}{\delta(\omega-\omega_d)(\delta-\omega+\omega_d)(\omega+\omega_d)(\delta+\omega+\omega_d)}\nonumber\\
    &\approx -2\imath \hat{a}e^{-\imath\phi}\frac{\omega}{\omega_d^2-\omega^2}\frac{e^{\imath\delta t_o}-e^{\imath\delta t}}{\delta}
\end{align}

Thus, 
\begin{align}
    \alpha = \imath\frac{\Omega_d\Omega_d}{\delta} \frac{\omega}{\omega_d^2-\omega^2} e^{-\imath\phi}(e^{\imath\delta t_o}-e^{\imath\delta t}) = \imath \Omega_{\textrm{eff}} e^{-\imath\phi}\frac{e^{\imath\delta t_o}-e^{\imath\delta t}}{\delta},
\end{align}
where we've defined the effective Rabi frequency as $\Omega_{\textrm{eff}} = \Omega_d\Omega_d \omega/(\omega_d^2-\omega^2)$.

\section{Appendix B: Displacement due to two motional Raman transitions}

To understand the process, suppose that two electric fields of equal strength are present at frequencies $\omega_1$ and $\omega_2$ in the dipole configuration.
We define two parameters $\omega_\mu=(\omega_1+\omega_2)/2$ and $\Delta\omega=(\omega_2-\omega_1)$.
We then apply a quadrupole tome at $\omega_q = \omega_\mu + \omega_o + \delta$.
The resulting displacement, ignoring any intermodulation between the two dipole frequencies, can be expressed as: 
\begin{align}
    \alpha =\sum_{i=1}^2\imath \Omega_{\textrm{eff}} e^{-\imath\phi_i}\frac{e^{\imath\delta_i t_o}-e^{\imath\delta_i (T+t_o)}}{\delta} = \sum_{i=1}^2\imath \Omega_{\textrm{eff}} e^{\imath\delta_i t_o} e^{-\imath\phi_i}\frac{1-e^{\imath\delta_i T}}{\delta}. 
\end{align}
The dipole tones are at $\omega_1 = \omega_\mu - \Delta\omega/2$ and $\omega_2 = \omega_\mu + \Delta\omega/2$. 
The detuning in our displacement operator is defined as $\delta = \omega_q -\omega_d-\omega$.
Thus, $\delta_1 = \omega_q-\omega_1-\omega = \delta - \Delta\omega/2$, while $\delta_2 = \delta + \Delta\omega/2-\omega$.
Thus, for a displacement starting at $t_o = 0$, we have:
\begin{align}
    \alpha = \imath\Omega_\textrm{eff} \left(e^{-\imath\phi_1}\frac{1-e^{i(\delta - \Delta\omega/2)T}}{\delta - \Delta\omega/2} + e^{-\imath\phi_2} \frac{1-e^{i(\delta+\Delta\omega/2)T}}{\delta+\Delta\omega/2}\right).
\end{align}
We have explicitly allowed for the two electric fields to be incoherent, i.e. $\phi_1 \neq \phi_2$.
We assume these phases jump shot-to-shot but are roughly stable during the time $T$.
Ignoring a global phase, the displacement becomes:
\begin{align}
    \alpha = \imath\Omega_\textrm{eff} \left(\frac{1-e^{i(\delta - \Delta\omega/2)T}}{\delta - \Delta\omega/2} + e^{-\imath\Delta\phi} \frac{1-e^{i(\delta+\Delta\omega/2)T}}{\delta+\Delta\omega/2}\right),
\end{align}
with $\Delta\phi = \phi_2-\phi_1$.

We assume that the two unknown tones are incoherent, i.e. $\Delta\phi$ is random but slowly varying over the course of a measurement trial.
(If the two tones are coherent, i.e. $\Delta\phi$ is fixed, then super-resolution can also be achieved by the means outlined here.)
Therefore, the state of the oscillator is described by the density matrix
\begin{align}
    \hat{\rho} = \frac{1}{2\pi}\int \ket{\alpha(\omega,\Delta\phi)}\bra{\alpha(\omega,\Delta\phi)}d(\Delta\phi) = \frac{1}{2}\left(\ket{\alpha(\omega,0)}\bra{\alpha(\omega,0)} + \ket{\alpha(\omega,\pi)}\bra{\alpha(\omega,\pi)}\right),
\end{align}
and a measurement of the phonon number returns $\langle n\rangle =(|\alpha(\omega,0)|^2 + |\alpha(\omega,\pi)|^2)/2$.

To effect super-resolution, we apply a WMDD sequence, where after time $T/2$ we shift the phase of the quadrupole tone by $\pi$ and evolve for another time of $T/2$. 
The total displacement is:
\begin{align}
    \alpha &= \imath\Omega_\textrm{eff} \left(\frac{1-e^{\imath(\delta - \Delta\omega/2)\frac{T}{2}}}{\delta - \Delta\omega/2} + e^{-\imath\Delta\phi} \frac{1-e^{i(\delta+\Delta\omega/2)\frac{T}{2}}}{\delta+\Delta\omega/2}\right)\\ &+ \imath e^{\imath\pi} \Omega_\textrm{eff} \left(e^{\imath(\delta-\Delta\omega/2)\frac{T}{2}} \frac{1-e^{\imath(\delta - \Delta\omega/2)\frac{T}{2}}}{\delta - \delta\omega} + e^{\imath(\delta+\Delta\omega/2)\frac{T}{2}} e^{-\imath\Delta\phi} \frac{1-e^{\imath(\delta+\Delta\omega/2)\frac{T}{2}}}{\delta+\Delta\omega/2}\right)\nonumber\\
    &=\imath\Omega_\textrm{eff} \left(\frac{\left(1-e^{\imath(\delta - \Delta\omega/2)\frac{T}{2}}\right)^2}{\delta - \Delta\omega/2} + e^{-\imath\Delta\phi}\frac{\left(1-e^{\imath(\delta+\Delta\omega/2)\frac{T}{2}}\right)^2}{\delta+\Delta\omega/2}\right)
\end{align}
And for $\delta = 0$ -- i.e. $\omega_q = \omega_\mu + \omega$, we have:
\begin{align}
    \alpha  &=\imath\Omega_\textrm{eff} \left(e^{-\imath\Delta\phi}\frac{\left(1-e^{\imath \Delta\omega\frac{T}{4}}\right)^2}{\Delta\omega/2} -\frac{\left(1-e^{-\imath \Delta\omega\frac{T}{4}}\right)^2}{\Delta\omega/2}\right).
\end{align}
And the amplitude of the displacement is:
\begin{align}
    |\alpha|^2 &= 256 \frac{\Omega_\textrm{eff}^2}{\Delta\omega^2} \sin^4\left(\frac{\Delta\omega T}{8}\right) \sin^2\left(\frac{\Delta\omega T/2-\Delta\phi}{2}\right)
\end{align}
For super resolution we are in the limit $\Delta\omega T\ll 1$,
for $\Delta\phi = 0$, we have:
\begin{align}
    |\alpha|^2 &\approx \frac{ \Omega_\textrm{eff}}{256}^2  \Delta\omega^4 T^6,
\end{align}
while for $\Delta\phi = \pi$, we have:
\begin{align}
    |\alpha|^2 &\approx  \frac{\Omega_\textrm{eff}^2}{16} \Delta\omega^2 T^4 .
\end{align}

And for a random phase, we have:

\onecolumngrid
\begin{align}
    \langle |\alpha|^2\rangle_{\Delta\phi} &= \frac{1}{2\pi}\int_0^{2\pi} 256 \frac{\Omega_\textrm{eff}^2}{\Delta\omega^2} \sin^4\left(\frac{\Delta\omega T}{8}\right) \sin^2\left(\frac{\Delta\omega T/2-\Delta\phi}{2}\right) d\Delta\phi = 128 \frac{\Omega_\textrm{eff}^2}{\Delta\omega^2} \sin^4\left(\frac{\Delta\omega T}{8}\right).
\end{align}

\subsection{Appendix C: Optimizing the Probe Frequency $\omega_p$}
\begin{figure}
    \centering    \includegraphics[width=0.5\textwidth]{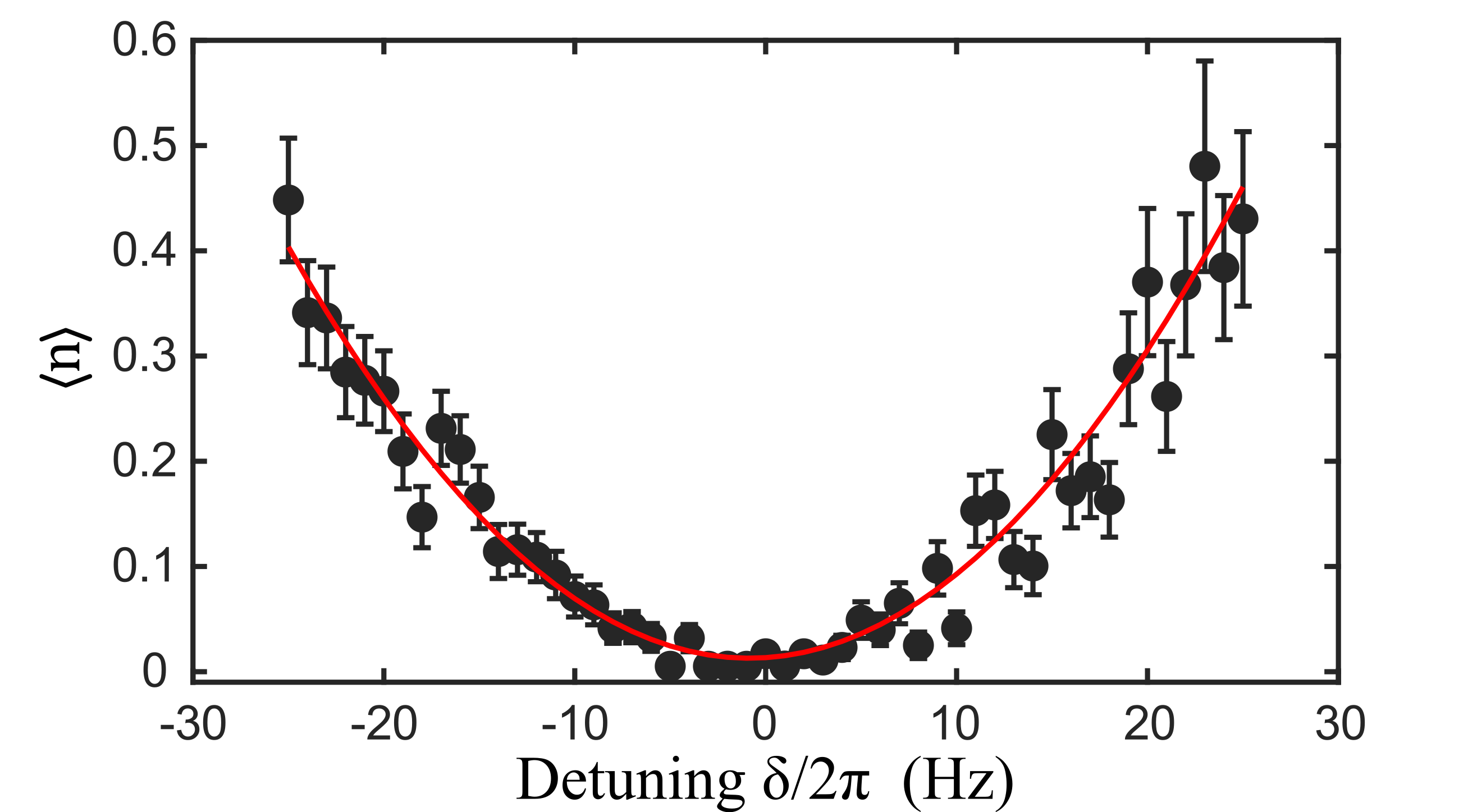}
    \caption{
    Optimizing the probe frequency $\omega_p$ for the super-resolution protocol.
    The mean phonon number $\langle n \rangle$ is measured while scanning the probe detuning $\delta$ as the WMDD protocol is applied.
    The optimal probe frequency $\omega_{p}$ is determined by minimizing $\langle n \rangle$.
    The solid line is theoretical fit to the data, which gives an uncertainty $\sigma_{\omega_p}\approx 2\pi \cdot500 \textrm{ mHz}$.
    }
    \label{fig:X1_peak}
\end{figure}

The probe frequency $\omega_p$ is optimized by scanning the probe detuning $\delta$ as the WMDD protocol is applied.
$\omega_{p}$ is optimized when the measured $\langle n \rangle$ is minimized, as shown in Fig.~\ref{fig:X1_peak}.
The theoretical fit estimates an uncertainty $\sigma_{\omega_p}\approx 2\pi \cdot500 \textrm{ mHz}$.

However, suppose the applied $\omega_p$ deviates from the optimal value by some frequency offset $\omega_{\textrm{offset}}$.
The measured mean phonon number $\langle n \rangle$ is then:
\begin{align}
    \langle n \rangle = \frac{\Omega_\textrm{eff}^2}{8} \bigg(T^4 \Delta\omega^2/4 &+ \nonumber \
    T^4 \omega_{\textrm{offset}}^2 - T^6 \Delta\omega^2 \omega_{\textrm{offset}}^2/4 \bigg)
    \label{eq:wp_n_deviation}
\end{align}
For small $\omega_{\textrm{offset}}$, i.e. $\omega_{\textrm{offset}}^2T^2 \ll 1$, this expression simplifies to $\langle n \rangle=\frac{\Omega_\textrm{eff}^2}{8} \left(T^4 \Delta\omega^2/4 + T^4 \omega_{\textrm{offset}}^2 \right)$.

Compared to the standard expression (Eq.~\eqref{eq:incoherent}), this frequency error introduces an additional offset independent of $\Delta\omega$. 
To correct for this offset, a two-point measurement can be performed, measuring $\langle n \rangle$ using our protocol with shifting $\omega_{\textrm{offset}}$ by a known $\Delta\omega$.
By jointly analyzing the two measurements, nonzero $\omega_{\textrm{offset}}$ can be compensated.
In our experiment, the value of $\omega_{\textrm{offset}}$ can be controlled determined to $\approx 2\pi\cdot500 \textrm{ mHz}$, as demonstrated in Fig.~\ref{fig:X1_peak}.
However, the presence of a nonzero offset increases the uncertainty in $\langle n \rangle$ during an Allan deviation, artificially enlarging $\Delta\omega$ and degrading the measurement resolution.

\subsection{Appendix D: Phase-Averaged Estimator For Incoherent Signals (when the relative phase $\Delta\phi$ varies between shots)}
The inversion estimator is employed to determine the mean phonon number $\langle n \rangle$ \cite{Wu2025}. 
Specifically, for a coherent state $\ket{\alpha}$, $\langle n \rangle$ can be inferred from the ratio of the motional sideband excitation probabilities $P_r$ and $P_b$.
Here, $P_{r/b}$ is the single-shot probability of a dark-state measurement following a red (blue) sideband Rabi pulse for duration $t_{\pi} = \pi / \Omega_{01}$, where $\Omega_{01}$ is the Rabi frequency of the $\ket{^{2}S_{1/2},n=0} \leftrightarrow \ket{^{2}D_{5/2},n=1}$ transition. See Ref.~\cite{Wu2025} for further details.

\begin{figure}
    \centering
    \includegraphics[width=0.5\textwidth]{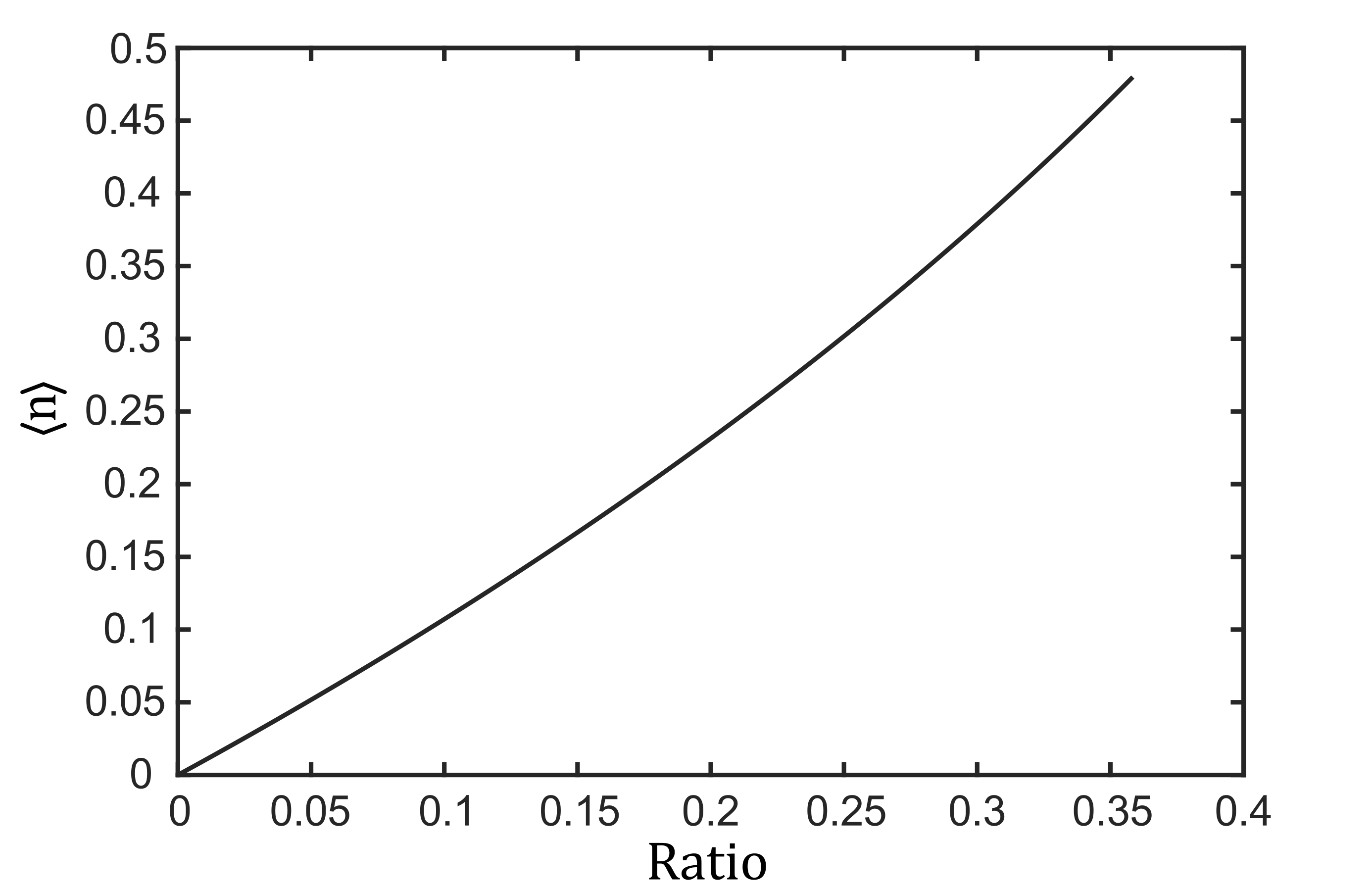}
    \caption{
    The conversion function $g\left(x\right)$ used to convert from a sideband ratio measurement to a mean phonon number $\langle n \rangle$ when the relative phase $\Delta \phi$ varies randomly between measurements.
    }
    \label{fig:conversion}
\end{figure}

To simulate incoherence between the signal tones, their relative phase $\Delta\phi$ is randomly drawn from a uniform distribution over $\left[0, 2\pi \right]$ for each measurement.
Despite this phase randomization, precise estimation of $\langle n \rangle$ remains possible.
The estimator can be adapted by instead first averaging the detection probabilities over all repetitions before applying the inversion estimator.
This phase-averaged estimator for $\langle n \rangle$ is thus:
\begin{align}
    \check{n} = g \left (\frac{\langle P_r \rangle|_{\Delta\phi}}{\langle P_b \rangle|_{\Delta\phi}} \right) = g \left( \frac{k/N}{j/M} \right),
\end{align}
where $k$ ($j$) is the number of dark-state measurements observed in $N$ ($M$) trials after application of a Rabi pulse on the red (blue) motional sideband.
The conversion function $g(x)$ from the averaged sideband ratio to $\langle n \rangle$ is shown in Fig.~\ref{fig:conversion}.

\subsection{Appendix E: Characterizing the Effective Rabi Frequency $\Omega_\textrm{eff}$}
\begin{figure}
    \centering
    \includegraphics[width = 0.5\textwidth]{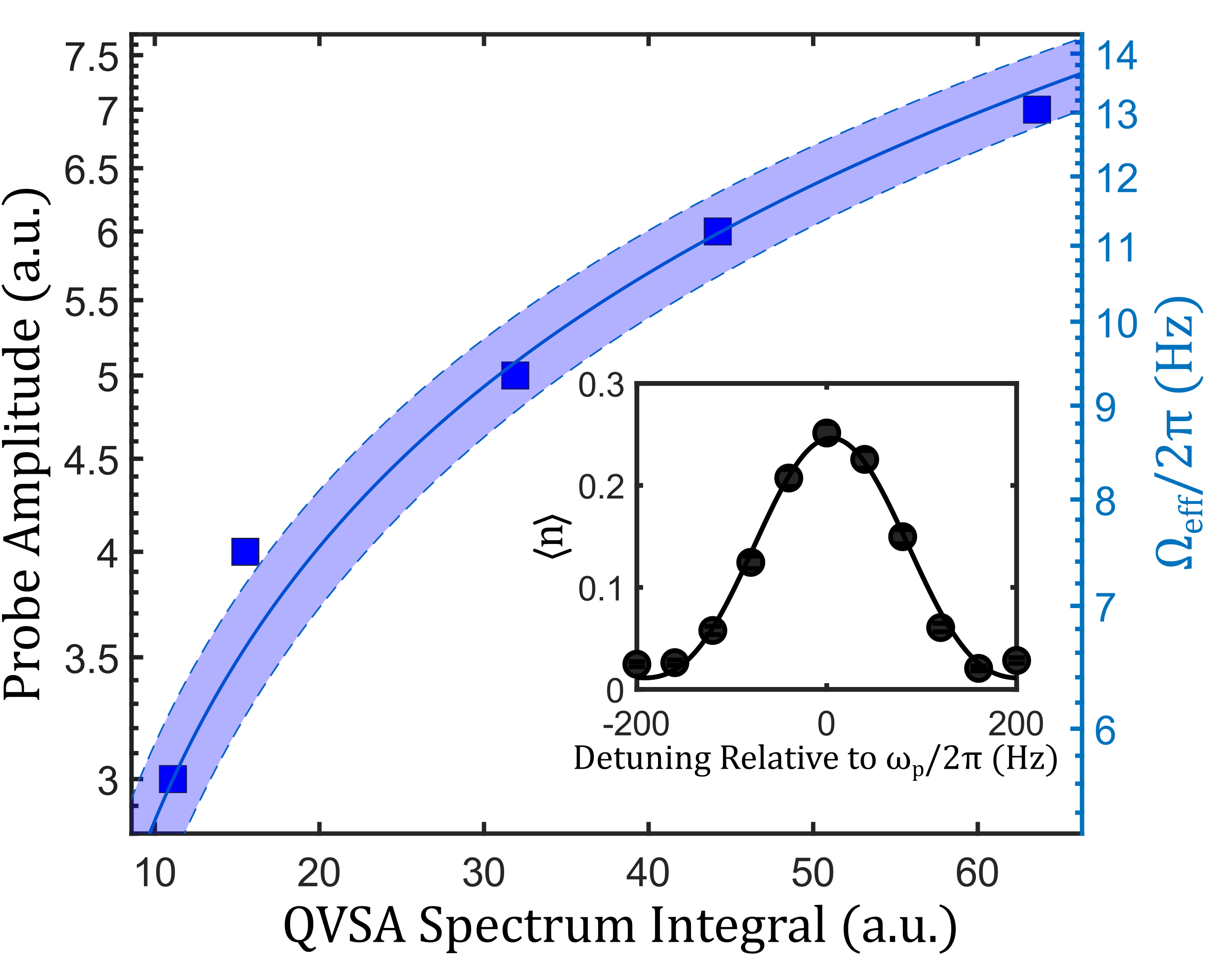}
    \caption{ 
    Robust calibration of the effective Rabi frequency $\Omega_{\textrm{eff}}$.
    $\Omega_{\textrm{eff}}$ is measured by integrating the total lineshape, obtained via conventional spectroscopy.
    Data (blue squared) are fitted to a function (blue line) $\textrm{Area} \approx  1.8 \ T\Omega_{\textrm{eff}}^2$.
    The shaded band indicates the 68\% prediction band, which corresponds to a $4\%$ prediction error in $\Omega_{\textrm{eff}}$.
    The inset displays a sample spectrum used to calculate an individual data point, here shown for an interrogation time $T = 5 \textrm{ ms}$.
    Error bars in the inset are too small to be observed.
    }
    \label{fig:rabi}
\end{figure}

While coarse characterization of the power (or amplitude) of the input signals is achievable with standard methods (e.g. a spectrum analyzer), we instead employ the ion itself as a \textit{quantum} signal analyzer to effect a more accurate \textit{in situ} calibration~\cite{Wu2025}. 
We consider the case of two incoherent signal tones with equal, but unknown, amplitudes.
We outline two approaches to determine the resulting effective Rabi frequency $\Omega_{\textrm{eff}}$.
The straightforward approach is to simply measure the peak mean phonon number $\langle n \rangle$ of the spectroscopic lineshape with the probe frequency set to either $\omega_{\mu} \pm \omega_{o}$.
However, this is method susceptible to drifts in $\omega_{o}$.
To mitigate these errors, a more robust method extracts $\Omega_{\textrm{eff}}$ by integrating the total lineshape, illustrated in the inset of Fig.~\ref{fig:rabi}.
The shaded band represents the $68\%$ prediction band.
A function
\begin{align}
    \textrm{Area} &= \Omega_{\textrm{eff}}^2\int_{-1/T}^{1/T}{\langle \left|\alpha\right|^2 \rangle_{\Delta\phi}} \ df_p \\
    &\approx  1.8 \ T \Omega_{\textrm{eff}}^2 \nonumber
\end{align}
is used for the fit, where the interrogation time $T$ is set directly in hardware, and $f_p=\frac{\omega_p}{2\pi}$.
This expression is accurate to within $2\%$ as $\frac{T \ \Delta\omega}{2\pi} \le 0.4$.
$\Omega_{\textrm{eff}}$, as used in Fig.\ref{fig:X2}, is then extracted via a calibration curve based on measured phonon, shown in Fig.~\ref{fig:rabi}.
An interrogation time $T = 5 \textrm{ ms}$ is used.
The resulting estimate of $\Omega_{\textrm{eff}}$ has a standard error of approximately $4\%$, as defined by the measured prediction band.

\subsection{Appendix F: Compensation of Motional Background and Drifts Between Measurements}
To compensate for the drifts in the motional background between experimental runs, a coarse calibration (shown in Fig.~\ref{fig:allan}(a)) is performed before each measurement of $\Delta \omega$.
We then measure $\langle n \rangle$ at different $\Delta \omega$ with the WMDD sequence, using the same known Rabi frequency $\Omega_{\textrm{eff}}$.

Further background suppression is achieved by conducting concurrent calibrations, i.e. measuring $\langle n \rangle$ at some known $\Delta\omega$.
Motional background from e.g. system drifts can then be suppressed by subtracting the background, given by the calibration measurements, from the target measurement, prior to processing via Eq.~\eqref{eq:incoherent}.

\subsection{Appendix G: Accounting for Amplitude Imbalance}
When the amplitudes of the two signal tones are unequal, the spectroscopic peak shifts from the midpoint  $\frac{\omega_1+\omega_2}{2}$ to a weighted average:
\begin{align}
    \omega_{\textrm{peak}}=\frac{\Omega_1\omega_1+\Omega_2\omega_2}{\Omega_1+\Omega_2},
    \label{eq:peak}
\end{align}
where $\Omega_1$ and $\Omega_2$ are the respective effective Rabi frequencies.
The integrated lineshape is thus:
\begin{align}
    \textrm{Area} \approx 0.9 T \left( \Omega_1^2 + \Omega_2^2 \right).
    \label{eq:area}
\end{align}
Similarly, the mean phonon number when the relative phase $\Delta\phi$ is randomized is:
\begin{align}
    \langle \left|\alpha\right|^2\rangle_{\Delta\phi} \approx  \frac{1}{8} \left(\frac{\Omega_1 \Omega_2}{\Omega_1+ \Omega_2} \right)^2 \Delta\omega^2 \ T^4.
    \label{eq:uneven}
\end{align}
Measurement of the ratio $R=\Omega_{1} / \Omega_{2}$ is thus required to determine $\Delta\omega$.
Three approaches can be used to determine the amplitude ratio $R$.
The first is a classical, self-homodyne technique that uses typical laboratory test and measurement equipment (e.g. an oscilloscope, a spectrum analyzer).
The input signals can be expressed:
\begin{align}
    f{\left(t\right)} = a_{1} \cos{\left(\omega_1 t \right)} + a_{2} \cos{\left(\omega_2 t \right)}
    \label{eq:input}
\end{align}
The inputs are split with equal amplitude using an RF splitter and fed into the LO and RF ports of a mixer, producing an output
\begin{align}
    f{\left( t \right)} &= \frac{a_1^2 + a_2^2}{2} +\frac{a_1^2}{2}\cos{\left(2\omega_1t \right)} + \frac{a_2^2}{2}\cos{\left(2\omega_2t \right)}\\ \nonumber
    & + a_1 a_2 \left( \cos{\left((\omega_1-\omega_2)t \right)} + \cos{\left((\omega_1+\omega_2)t \right)} \right).
    \label{eq:input}
\end{align}
The DC component $\frac{a_1^2 + a_2^2}{2}$ can be isolated by applying a low-pass filter to the output and measuring on e.g. an oscilloscope.
Similarly, in the super-resolution regime where $2\omega_1 \approx 2 \omega_2 \approx \omega_1 + \omega_2$, the $a_1^2 + a_2^2 + 2 a_1 a_2$ component can be measured on e.g. a spectrum analyzer.
Together, these measurements can be processed to yield the desired ratio $R = \frac{a_1}{a_2}=\frac{\Omega_1}{\Omega_2}$.

The second approach combines quantum (ion-based) measurements with classical techniques.
A key challenge with quantum systems is that the Schrödinger equation is inherently linear, making it difficult to obtain nonlinear cross-terms, e.g. $a_{1} a_{2}$.
The requisite nonlinearity can instead be introduced classically, by e.g. applying the inputs to an RF mixer, as in the classical method.
However, instead of using traditional test and measurement equipment for detection, the mixed outputs can be detected using the quantum system at hand; here, the trapped ion.
Detection of the mixed signal using the technique in Fig.~\ref{fig:rabi} now yields $\Omega_1^2+\Omega_2^2+2\Omega_1\Omega_2$, which can be combined with the direct measurement of Eq.~\eqref{eq:uneven} to yield the amplitude ratio $R$.
Finally, the amplitude ratio can be detected using only the QHO system at hand.
Excitation of a nonlinear, second-order subharmonic generates an intermodulation term proportional to $\Omega_1 \Omega_2$, effectively functioning like the RF mixer in the classical method~\cite{WuH2025}.
Specifically, integration of the lineshape obtained by a scan of $\omega_p$ about the second-order subharmonic gives:
\begin{align}
    \textrm{Area}_2 \propto \Omega_1^2 + \Omega_2^2 + 2\Omega_1\Omega_2,
\end{align}
where proportionality factors can be derived from the relevant theory. This complements the measurement Eq.~\eqref{eq:peak} to allow exact calculation of the amplitude ratio $R$.
\subsection{Appendix H: Quantum Fisher Information under the WMDD Protocol}
The density matrix of the system under the WMDD protocol is:
\begin{align}
    \hat{\rho} &= \frac{1}{2\pi}\int \ket{\alpha_{SR}(\Delta \omega,\Delta\phi, T)}\bra{\alpha_{SR}(\Delta \omega,\Delta\phi, T)}d(\Delta\phi).
\end{align}
From convexity of the QFI \cite{wolf2015}, we obtain the inequality:
\begin{align}
    \mathcal{F}_Q(\rho)\leq \frac{1}{2\pi}\int{\mathcal{F}_Q \left(\rho\left(\Delta\phi\right)\right) d\Delta\phi}
    \label{eq:qfi_convexity}
\end{align}

Each pure state $\rho\left(\Delta\phi \right)$ corresponds to a displaced vacuum state of the form $\ket{\alpha_{\textrm{SR}} \left(\Delta \omega,\Delta\phi, T\right)} = e^{i\alpha_{\textrm{SR}} \left( \hat{a} + \hat{a}^{\dagger} \right)}\ket{0}$, where the displacement is:
\begin{align}
    \langle \alpha_{\textrm{SR}} \rangle = \frac{16\Omega_\textrm{eff}}{\Delta\omega} \sin^{2}{\left( \frac{\Delta\omega T}{8} \right)} \sin{\left(\frac{\Delta\omega T/2-\Delta\phi}{2}\right)}.
    \label{eq:phase2}
\end{align}
In the super-resolved limit $\Delta\omega T \ll1$, this simplifies to
$\langle \alpha_{\textrm{SR}} \rangle = \frac{\Omega_\textrm{eff}T^2 \Delta\omega}{4} \sin{\left(\frac{\Delta\omega T/2-\Delta\phi}{2}\right)}$.

For a pure state of the form $\ket{\chi{ \left(\theta\right)}} = e^{i G\theta}\ket{\chi{\left(0\right)}}$,
the QFI is $\mathcal{F}{ \left(\theta\right)} = 4 \left( \langle G^2 \rangle - \langle G \rangle^2 \right)$.
We then find $\mathcal{F}\left( \rho\left(\Delta\phi\right) \right) = \frac{\Omega_\textrm{eff}^2 T^4}{4} \sin^2{\left(\frac{\Delta\omega T/2-\Delta\phi}{2}\right)}$.
Averaging over $\Delta\phi$, Eq.~\eqref{eq:qfi_convexity} becomes $\mathcal{F}_Q{\left( \rho \right)} \leq \frac{\Omega_\textrm{eff}^2 T^4}{8}$.
The WMDD limit in Fig.~\ref{fig:X2}(c) is based on this expression.

\subsection{Appendix I: Histogram of Super-resolution}
\begin{figure*}
    \centering
    \includegraphics[width =0.8\textwidth]{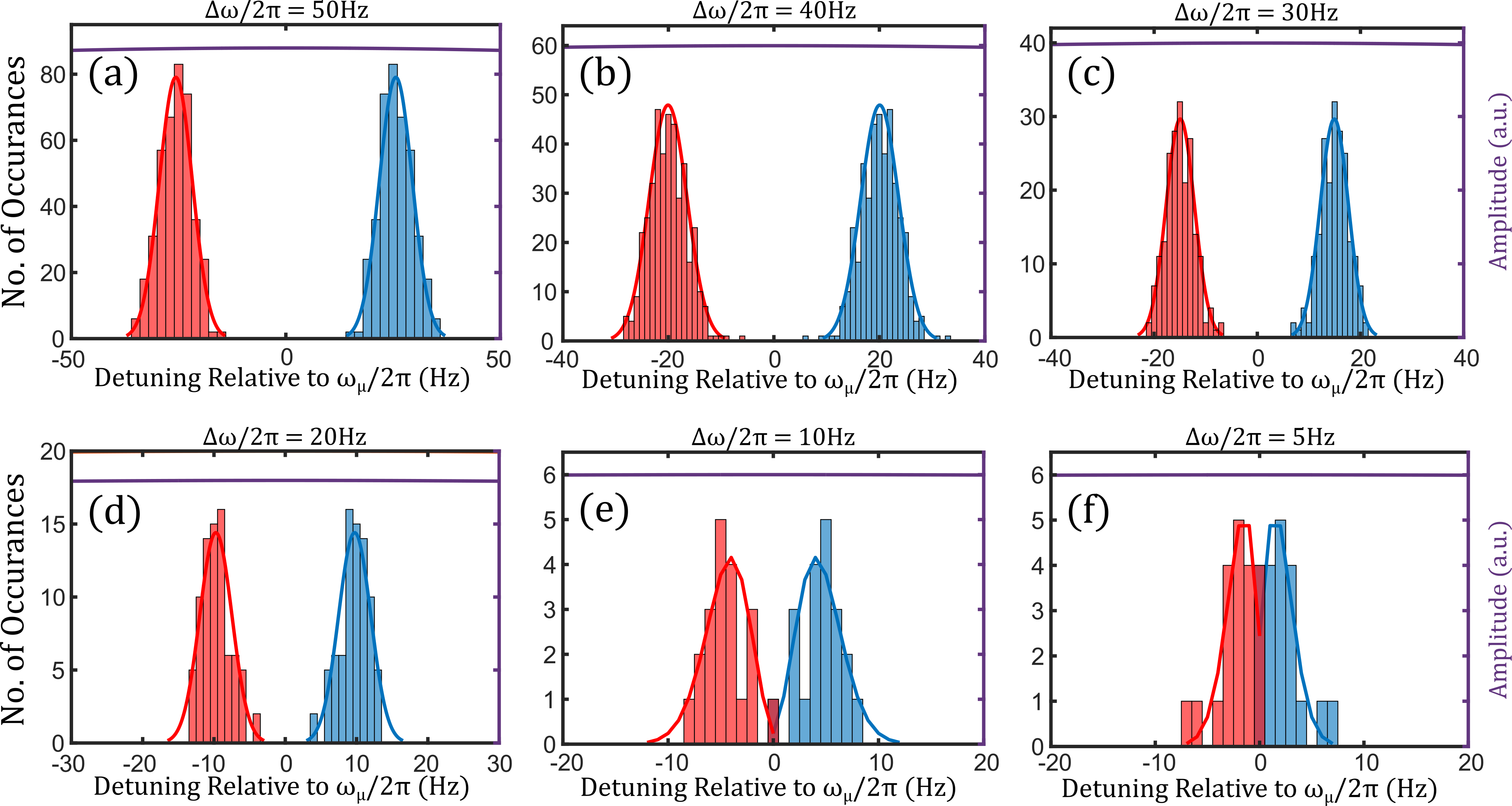}
    \caption{
    The summary of histogram of super-resolution for different $\Delta\omega$ measurement.
    The purple solid line is the result of conventional QVSA measurement. 
    The red/blue solid lines are fits to the data assuming a Poisson distribution.
    Each subfigure is averaged over M repetitions.
    (a-b) M=51; (c) M=102; (d) M=255; (e) M=510; (f) M=1020
    }
    \label{fig:hist}
\end{figure*}

Fig.~\ref{fig:hist} presents histograms of the super-resolution spectra for various values of $\Delta \omega$. The blue (red) distributions correspond to $\omega_{\mu} + \Delta \omega/2$ ($\omega_{\mu} - \Delta \omega/2$), with fits to the respective data assuming a Poisson distribution.
Our WMDD protocol readily resolves frequency separations as small as $\Delta\omega/2\pi = 5 \textrm{ Hz}$, a level of precision unattainable by conventional spectroscopy, illustrated by the solid purple line.

\subsection{Appendix J: Super-Resolution With Coherent Tones}
Consider two coherent signal tones separated in frequency by $\Delta\omega = 1 \textrm{ Hz}$.
After a duration of 0.5 seconds, the tones will accumulates a relative phase shift of $\pi$.
Adjustment of the dead time between measurements thus allows control over the relative phase between the tones, since is thus $\Delta\phi = \tau_{\textrm{dead}} \Delta\omega$.
Randomizing the dead time uniformly over $\left[ 0 \textrm{s}, 0.5\textrm{s} \right]$ is thus equivalent to randomizing the phase over $\left[ 0, \pi \right]$ between successive measurements, providing a means to engineer signal decoherence and enhance spectral resolution of coherent tones.

\subsection{Appendix K: Error Budget for the $\Delta \omega$ Estimation in Fig.~\ref{fig:X2}}
The total uncertainty in $\Delta\omega$ from all relevant parameters is given by:
\begin{align}
    \sigma_{\Delta\omega} = \Delta\omega\sqrt{\left( \frac{\sigma_{n}}{2n} \right)^2 + \left( \frac{\sigma_{ \Omega_{\textrm{eff}}}}{\Omega_{\textrm{eff}}} \right)^2 + \left( \frac{2\sigma_{ 
     T}}{T} \right)^2}.
\end{align}
Here, the uncertainty in the phonon number $\sigma_n$ is evaluated by conducting an overlapping Allan deviation on the dataset shown in Fig.~\ref{fig:allan}(b–d).
As discussed previously, the uncertainty of $\omega_p$ leads into the enhanced uncertainty in the phonon population and has been incorporated into $\sigma_n$ from the Allan deviation analysis.
$T$ can be accurately controlled in hardware (e.g. artiq).
The error values corresponding to Fig.~\ref{fig:X2}(b-c) are summarized in Table~\ref{tab:error_budget}.
\begin{table*}
    \caption{Error budget in $\Delta\omega$ estimation in Fig.~\ref{fig:X2}}
    \centering
    \begin{tabular}{|c|c|c|c|c|c|c|c|}
        \toprule
        $\sigma_n/n$(2.5Hz) & $\sigma_n/n$ (5Hz)& $\sigma_n/n$ (10Hz) &$\sigma_n/n$(15Hz) & $\sigma_n/n$ (20Hz)& $\sigma_n/n $(25Hz)& $\sigma_{\Omega_{\textrm{eff}}}/\Omega_{\textrm{eff}}$ & $\sigma_T/T$ \\
        \midrule
        0.67& 0.42 & 0.13&0.06& 0.06 & 0.06& 0.04& 1e-4 \\
        \bottomrule
    \end{tabular}
    \label{tab:error_budget}
\end{table*}

%% file: main_text.bbl
\begin{thebibliography}{37}%
\makeatletter
\providecommand \@ifxundefined [1]{%
 \@ifx{#1\undefined}
}%
\providecommand \@ifnum [1]{%
 \ifnum #1\expandafter \@firstoftwo
 \else \expandafter \@secondoftwo
 \fi
}%
\providecommand \@ifx [1]{%
 \ifx #1\expandafter \@firstoftwo
 \else \expandafter \@secondoftwo
 \fi
}%
\providecommand \natexlab [1]{#1}%
\providecommand \enquote  [1]{``#1''}%
\providecommand \bibnamefont  [1]{#1}%
\providecommand \bibfnamefont [1]{#1}%
\providecommand \citenamefont [1]{#1}%
\providecommand \href@noop [0]{\@secondoftwo}%
\providecommand \href [0]{\begingroup \@sanitize@url \@href}%
\providecommand \@href[1]{\@@startlink{#1}\@@href}%
\providecommand \@@href[1]{\endgroup#1\@@endlink}%
\providecommand \@sanitize@url [0]{\catcode `\\12\catcode `\$12\catcode `\&12\catcode `\#12\catcode `\^12\catcode `\_12\catcode `\%12\relax}%
\providecommand \@@startlink[1]{}%
\providecommand \@@endlink[0]{}%
\providecommand \url  [0]{\begingroup\@sanitize@url \@url }%
\providecommand \@url [1]{\endgroup\@href {#1}{\urlprefix }}%
\providecommand \urlprefix  [0]{URL }%
\providecommand \Eprint [0]{\href }%
\providecommand \doibase [0]{https://doi.org/}%
\providecommand \selectlanguage [0]{\@gobble}%
\providecommand \bibinfo  [0]{\@secondoftwo}%
\providecommand \bibfield  [0]{\@secondoftwo}%
\providecommand \translation [1]{[#1]}%
\providecommand \BibitemOpen [0]{}%
\providecommand \bibitemStop [0]{}%
\providecommand \bibitemNoStop [0]{.\EOS\space}%
\providecommand \EOS [0]{\spacefactor3000\relax}%
\providecommand \BibitemShut  [1]{\csname bibitem#1\endcsname}%
\let\auto@bib@innerbib\@empty
\bibitem [{\citenamefont {Helstrom}(1969)}]{Helstrom1969}%
  \BibitemOpen
  \bibfield  {author} {\bibinfo {author} {\bibfnamefont {C.~W.}\ \bibnamefont {Helstrom}},\ }\bibfield  {title} {\bibinfo {title} {Quantum detection and estimation theory},\ }\href {https://doi.org/10.1007/BF01007479} {\bibfield  {journal} {\bibinfo  {journal} {Journal of Statistical Physics}\ }\textbf {\bibinfo {volume} {1}},\ \bibinfo {pages} {231} (\bibinfo {year} {1969})}\BibitemShut {NoStop}%
\bibitem [{\citenamefont {Wootters}(1981)}]{Wootters1981}%
  \BibitemOpen
  \bibfield  {author} {\bibinfo {author} {\bibfnamefont {W.~K.}\ \bibnamefont {Wootters}},\ }\bibfield  {title} {\bibinfo {title} {Statistical distance and hilbert space},\ }\href {https://doi.org/10.1103/PhysRevD.23.357} {\bibfield  {journal} {\bibinfo  {journal} {Phys. Rev. D}\ }\textbf {\bibinfo {volume} {23}},\ \bibinfo {pages} {357} (\bibinfo {year} {1981})}\BibitemShut {NoStop}%
\bibitem [{\citenamefont {Braunstein}\ and\ \citenamefont {Caves}(1994)}]{Braunstein1994}%
  \BibitemOpen
  \bibfield  {author} {\bibinfo {author} {\bibfnamefont {S.~L.}\ \bibnamefont {Braunstein}}\ and\ \bibinfo {author} {\bibfnamefont {C.~M.}\ \bibnamefont {Caves}},\ }\bibfield  {title} {\bibinfo {title} {Statistical distance and the geometry of quantum states},\ }\href {https://doi.org/10.1103/PhysRevLett.72.3439} {\bibfield  {journal} {\bibinfo  {journal} {Phys. Rev. Lett.}\ }\textbf {\bibinfo {volume} {72}},\ \bibinfo {pages} {3439} (\bibinfo {year} {1994})}\BibitemShut {NoStop}%
\bibitem [{\citenamefont {Degen}\ \emph {et~al.}(2017)\citenamefont {Degen}, \citenamefont {Reinhard},\ and\ \citenamefont {Cappellaro}}]{Degen2017}%
  \BibitemOpen
  \bibfield  {author} {\bibinfo {author} {\bibfnamefont {C.~L.}\ \bibnamefont {Degen}}, \bibinfo {author} {\bibfnamefont {F.}~\bibnamefont {Reinhard}},\ and\ \bibinfo {author} {\bibfnamefont {P.}~\bibnamefont {Cappellaro}},\ }\bibfield  {title} {\bibinfo {title} {Quantum sensing},\ }\href {https://doi.org/10.1103/RevModPhys.89.035002} {\bibfield  {journal} {\bibinfo  {journal} {Rev. Mod. Phys.}\ }\textbf {\bibinfo {volume} {89}},\ \bibinfo {pages} {035002} (\bibinfo {year} {2017})}\BibitemShut {NoStop}%
\bibitem [{\citenamefont {Pezz\`e}\ \emph {et~al.}(2018)\citenamefont {Pezz\`e}, \citenamefont {Smerzi}, \citenamefont {Oberthaler}, \citenamefont {Schmied},\ and\ \citenamefont {Treutlein}}]{Pezze2018}%
  \BibitemOpen
  \bibfield  {author} {\bibinfo {author} {\bibfnamefont {L.}~\bibnamefont {Pezz\`e}}, \bibinfo {author} {\bibfnamefont {A.}~\bibnamefont {Smerzi}}, \bibinfo {author} {\bibfnamefont {M.~K.}\ \bibnamefont {Oberthaler}}, \bibinfo {author} {\bibfnamefont {R.}~\bibnamefont {Schmied}},\ and\ \bibinfo {author} {\bibfnamefont {P.}~\bibnamefont {Treutlein}},\ }\bibfield  {title} {\bibinfo {title} {Quantum metrology with nonclassical states of atomic ensembles},\ }\href {https://doi.org/10.1103/RevModPhys.90.035005} {\bibfield  {journal} {\bibinfo  {journal} {Rev. Mod. Phys.}\ }\textbf {\bibinfo {volume} {90}},\ \bibinfo {pages} {035005} (\bibinfo {year} {2018})}\BibitemShut {NoStop}%
\bibitem [{\citenamefont {Bouchet}\ \emph {et~al.}(2021)\citenamefont {Bouchet}, \citenamefont {Rotter},\ and\ \citenamefont {Mosk}}]{Bouchet2021}%
  \BibitemOpen
  \bibfield  {author} {\bibinfo {author} {\bibfnamefont {D.}~\bibnamefont {Bouchet}}, \bibinfo {author} {\bibfnamefont {S.}~\bibnamefont {Rotter}},\ and\ \bibinfo {author} {\bibfnamefont {A.~P.}\ \bibnamefont {Mosk}},\ }\bibfield  {title} {\bibinfo {title} {Maximum information states for coherent scattering measurements},\ }\href {https://doi.org/10.1038/s41567-020-01137-4} {\bibfield  {journal} {\bibinfo  {journal} {Nature Physics}\ }\textbf {\bibinfo {volume} {17}},\ \bibinfo {pages} {564} (\bibinfo {year} {2021})}\BibitemShut {NoStop}%
\bibitem [{\citenamefont {Giovannetti}\ \emph {et~al.}(2004)\citenamefont {Giovannetti}, \citenamefont {Lloyd},\ and\ \citenamefont {Maccone}}]{Vittorio2004}%
  \BibitemOpen
  \bibfield  {author} {\bibinfo {author} {\bibfnamefont {V.}~\bibnamefont {Giovannetti}}, \bibinfo {author} {\bibfnamefont {S.}~\bibnamefont {Lloyd}},\ and\ \bibinfo {author} {\bibfnamefont {L.}~\bibnamefont {Maccone}},\ }\bibfield  {title} {\bibinfo {title} {Quantum-enhanced measurements: Beating the standard quantum limit},\ }\href {https://doi.org/10.1126/science.1104149} {\bibfield  {journal} {\bibinfo  {journal} {Science}\ }\textbf {\bibinfo {volume} {306}},\ \bibinfo {pages} {1330} (\bibinfo {year} {2004})},\ \Eprint {https://arxiv.org/abs/https://www.science.org/doi/pdf/10.1126/science.1104149} {https://www.science.org/doi/pdf/10.1126/science.1104149} \BibitemShut {NoStop}%
\bibitem [{\citenamefont {Tsang}\ \emph {et~al.}(2016)\citenamefont {Tsang}, \citenamefont {Nair},\ and\ \citenamefont {Lu}}]{Tsang2016}%
  \BibitemOpen
  \bibfield  {author} {\bibinfo {author} {\bibfnamefont {M.}~\bibnamefont {Tsang}}, \bibinfo {author} {\bibfnamefont {R.}~\bibnamefont {Nair}},\ and\ \bibinfo {author} {\bibfnamefont {X.-M.}\ \bibnamefont {Lu}},\ }\bibfield  {title} {\bibinfo {title} {Quantum theory of superresolution for two incoherent optical point sources},\ }\href {https://doi.org/10.1103/PhysRevX.6.031033} {\bibfield  {journal} {\bibinfo  {journal} {Phys. Rev. X}\ }\textbf {\bibinfo {volume} {6}},\ \bibinfo {pages} {031033} (\bibinfo {year} {2016})}\BibitemShut {NoStop}%
\bibitem [{\citenamefont {Pa\'{u}r}\ \emph {et~al.}(2016)\citenamefont {Pa\'{u}r}, \citenamefont {Stoklasa}, \citenamefont {Hradil}, \citenamefont {S\'{a}nchez-Soto},\ and\ \citenamefont {Rehacek}}]{Paur2016}%
  \BibitemOpen
  \bibfield  {author} {\bibinfo {author} {\bibfnamefont {M.}~\bibnamefont {Pa\'{u}r}}, \bibinfo {author} {\bibfnamefont {B.}~\bibnamefont {Stoklasa}}, \bibinfo {author} {\bibfnamefont {Z.}~\bibnamefont {Hradil}}, \bibinfo {author} {\bibfnamefont {L.~L.}\ \bibnamefont {S\'{a}nchez-Soto}},\ and\ \bibinfo {author} {\bibfnamefont {J.}~\bibnamefont {Rehacek}},\ }\bibfield  {title} {\bibinfo {title} {Achieving the ultimate optical resolution},\ }\href {https://doi.org/10.1364/OPTICA.3.001144} {\bibfield  {journal} {\bibinfo  {journal} {Optica}\ }\textbf {\bibinfo {volume} {3}},\ \bibinfo {pages} {1144} (\bibinfo {year} {2016})}\BibitemShut {NoStop}%
\bibitem [{\citenamefont {Tang}\ \emph {et~al.}(2016)\citenamefont {Tang}, \citenamefont {Durak},\ and\ \citenamefont {Ling}}]{Tang2016}%
  \BibitemOpen
  \bibfield  {author} {\bibinfo {author} {\bibfnamefont {Z.~S.}\ \bibnamefont {Tang}}, \bibinfo {author} {\bibfnamefont {K.}~\bibnamefont {Durak}},\ and\ \bibinfo {author} {\bibfnamefont {A.}~\bibnamefont {Ling}},\ }\bibfield  {title} {\bibinfo {title} {Fault-tolerant and finite-error localization for point emitters within the diffraction limit},\ }\href {https://doi.org/10.1364/OE.24.022004} {\bibfield  {journal} {\bibinfo  {journal} {Opt. Express}\ }\textbf {\bibinfo {volume} {24}},\ \bibinfo {pages} {22004} (\bibinfo {year} {2016})}\BibitemShut {NoStop}%
\bibitem [{\citenamefont {Yang}\ \emph {et~al.}(2016)\citenamefont {Yang}, \citenamefont {Tashchilina}, \citenamefont {Moiseev}, \citenamefont {Simon},\ and\ \citenamefont {Lvovsky}}]{Yang2016}%
  \BibitemOpen
  \bibfield  {author} {\bibinfo {author} {\bibfnamefont {F.}~\bibnamefont {Yang}}, \bibinfo {author} {\bibfnamefont {A.}~\bibnamefont {Tashchilina}}, \bibinfo {author} {\bibfnamefont {E.~S.}\ \bibnamefont {Moiseev}}, \bibinfo {author} {\bibfnamefont {C.}~\bibnamefont {Simon}},\ and\ \bibinfo {author} {\bibfnamefont {A.~I.}\ \bibnamefont {Lvovsky}},\ }\bibfield  {title} {\bibinfo {title} {Far-field linear optical superresolution via heterodyne detection in a higher-order local oscillator mode},\ }\href {https://doi.org/10.1364/OPTICA.3.001148} {\bibfield  {journal} {\bibinfo  {journal} {Optica}\ }\textbf {\bibinfo {volume} {3}},\ \bibinfo {pages} {1148} (\bibinfo {year} {2016})}\BibitemShut {NoStop}%
\bibitem [{\citenamefont {Tham}\ \emph {et~al.}(2017)\citenamefont {Tham}, \citenamefont {Ferretti},\ and\ \citenamefont {Steinberg}}]{Tham2017}%
  \BibitemOpen
  \bibfield  {author} {\bibinfo {author} {\bibfnamefont {W.-K.}\ \bibnamefont {Tham}}, \bibinfo {author} {\bibfnamefont {H.}~\bibnamefont {Ferretti}},\ and\ \bibinfo {author} {\bibfnamefont {A.~M.}\ \bibnamefont {Steinberg}},\ }\bibfield  {title} {\bibinfo {title} {Beating rayleigh's curse by imaging using phase information},\ }\href {https://doi.org/10.1103/PhysRevLett.118.070801} {\bibfield  {journal} {\bibinfo  {journal} {Phys. Rev. Lett.}\ }\textbf {\bibinfo {volume} {118}},\ \bibinfo {pages} {070801} (\bibinfo {year} {2017})}\BibitemShut {NoStop}%
\bibitem [{\citenamefont {Rotem}\ \emph {et~al.}(2019)\citenamefont {Rotem}, \citenamefont {Gefen}, \citenamefont {Oviedo-Casado}, \citenamefont {Prior}, \citenamefont {Schmitt}, \citenamefont {Burak}, \citenamefont {McGuiness}, \citenamefont {Jelezko},\ and\ \citenamefont {Retzker}}]{Rotem2019}%
  \BibitemOpen
  \bibfield  {author} {\bibinfo {author} {\bibfnamefont {A.}~\bibnamefont {Rotem}}, \bibinfo {author} {\bibfnamefont {T.}~\bibnamefont {Gefen}}, \bibinfo {author} {\bibfnamefont {S.}~\bibnamefont {Oviedo-Casado}}, \bibinfo {author} {\bibfnamefont {J.}~\bibnamefont {Prior}}, \bibinfo {author} {\bibfnamefont {S.}~\bibnamefont {Schmitt}}, \bibinfo {author} {\bibfnamefont {Y.}~\bibnamefont {Burak}}, \bibinfo {author} {\bibfnamefont {L.}~\bibnamefont {McGuiness}}, \bibinfo {author} {\bibfnamefont {F.}~\bibnamefont {Jelezko}},\ and\ \bibinfo {author} {\bibfnamefont {A.}~\bibnamefont {Retzker}},\ }\bibfield  {title} {\bibinfo {title} {Limits on spectral resolution measurements by quantum probes},\ }\href {https://doi.org/10.1103/PhysRevLett.122.060503} {\bibfield  {journal} {\bibinfo  {journal} {Phys. Rev. Lett.}\ }\textbf {\bibinfo {volume} {122}},\ \bibinfo {pages} {060503} (\bibinfo {year} {2019})}\BibitemShut {NoStop}%
\bibitem [{\citenamefont {Gefen}\ \emph {et~al.}(2019)\citenamefont {Gefen}, \citenamefont {Rotem},\ and\ \citenamefont {Retzker}}]{Gefen2019}%
  \BibitemOpen
  \bibfield  {author} {\bibinfo {author} {\bibfnamefont {T.}~\bibnamefont {Gefen}}, \bibinfo {author} {\bibfnamefont {A.}~\bibnamefont {Rotem}},\ and\ \bibinfo {author} {\bibfnamefont {A.}~\bibnamefont {Retzker}},\ }\bibfield  {title} {\bibinfo {title} {Overcoming resolution limits with quantum sensing},\ }\href {https://doi.org/10.1038/s41467-019-12817-y} {\bibfield  {journal} {\bibinfo  {journal} {Nature Communications}\ }\textbf {\bibinfo {volume} {10}},\ \bibinfo {pages} {4992} (\bibinfo {year} {2019})}\BibitemShut {NoStop}%
\bibitem [{\citenamefont {Wu}\ \emph {et~al.}(2025{\natexlab{a}})\citenamefont {Wu}, \citenamefont {Mitts}, \citenamefont {Ho}, \citenamefont {Rabinowitz},\ and\ \citenamefont {Hudson}}]{Wu2025}%
  \BibitemOpen
  \bibfield  {author} {\bibinfo {author} {\bibfnamefont {H.}~\bibnamefont {Wu}}, \bibinfo {author} {\bibfnamefont {G.~D.}\ \bibnamefont {Mitts}}, \bibinfo {author} {\bibfnamefont {C.~Z.~C.}\ \bibnamefont {Ho}}, \bibinfo {author} {\bibfnamefont {J.~A.}\ \bibnamefont {Rabinowitz}},\ and\ \bibinfo {author} {\bibfnamefont {E.~R.}\ \bibnamefont {Hudson}},\ }\bibfield  {title} {\bibinfo {title} {Wideband electric field quantum sensing via motional raman transitions},\ }\bibfield  {journal} {\bibinfo  {journal} {Nature Physics}\ }\href {https://doi.org/10.1038/s41567-024-02753-0} {10.1038/s41567-024-02753-0} (\bibinfo {year} {2025}{\natexlab{a}})\BibitemShut {NoStop}%
\bibitem [{\citenamefont {Hempel}\ \emph {et~al.}(2013)\citenamefont {Hempel}, \citenamefont {Lanyon}, \citenamefont {Jurcevic}, \citenamefont {Gerritsma}, \citenamefont {Blatt},\ and\ \citenamefont {Roos}}]{Hempel2013}%
  \BibitemOpen
  \bibfield  {author} {\bibinfo {author} {\bibfnamefont {C.}~\bibnamefont {Hempel}}, \bibinfo {author} {\bibfnamefont {B.~P.}\ \bibnamefont {Lanyon}}, \bibinfo {author} {\bibfnamefont {P.}~\bibnamefont {Jurcevic}}, \bibinfo {author} {\bibfnamefont {R.}~\bibnamefont {Gerritsma}}, \bibinfo {author} {\bibfnamefont {R.}~\bibnamefont {Blatt}},\ and\ \bibinfo {author} {\bibfnamefont {C.~F.}\ \bibnamefont {Roos}},\ }\bibfield  {title} {\bibinfo {title} {Entanglement-enhanced detection of single-photon scattering events},\ }\href {https://doi.org/10.1038/nphoton.2013.172} {\bibfield  {journal} {\bibinfo  {journal} {Nature Photonics}\ }\textbf {\bibinfo {volume} {7}},\ \bibinfo {pages} {630} (\bibinfo {year} {2013})}\BibitemShut {NoStop}%
\bibitem [{\citenamefont {Burd}\ \emph {et~al.}(2019)\citenamefont {Burd}, \citenamefont {Srinivas}, \citenamefont {Bollinger}, \citenamefont {Wilson}, \citenamefont {Wineland}, \citenamefont {Leibfried}, \citenamefont {Slichter},\ and\ \citenamefont {Allcock}}]{Burd2019}%
  \BibitemOpen
  \bibfield  {author} {\bibinfo {author} {\bibfnamefont {S.~C.}\ \bibnamefont {Burd}}, \bibinfo {author} {\bibfnamefont {R.}~\bibnamefont {Srinivas}}, \bibinfo {author} {\bibfnamefont {J.~J.}\ \bibnamefont {Bollinger}}, \bibinfo {author} {\bibfnamefont {A.~C.}\ \bibnamefont {Wilson}}, \bibinfo {author} {\bibfnamefont {D.~J.}\ \bibnamefont {Wineland}}, \bibinfo {author} {\bibfnamefont {D.}~\bibnamefont {Leibfried}}, \bibinfo {author} {\bibfnamefont {D.~H.}\ \bibnamefont {Slichter}},\ and\ \bibinfo {author} {\bibfnamefont {D.~T.~C.}\ \bibnamefont {Allcock}},\ }\bibfield  {title} {\bibinfo {title} {Quantum amplification of mechanical oscillator motion},\ }\href {https://doi.org/10.1126/science.aaw2884} {\bibfield  {journal} {\bibinfo  {journal} {Science}\ }\textbf {\bibinfo {volume} {364}},\ \bibinfo {pages} {1163} (\bibinfo {year} {2019})},\ \bibinfo {note} {doi: 10.1126/science.aaw2884}\BibitemShut {NoStop}%
\bibitem [{\citenamefont {Gilmore}\ \emph {et~al.}(2021)\citenamefont {Gilmore}, \citenamefont {Affolter}, \citenamefont {Lewis-Swan}, \citenamefont {Barberena}, \citenamefont {Jordan}, \citenamefont {Rey},\ and\ \citenamefont {Bollinger}}]{Gilmore2021}%
  \BibitemOpen
  \bibfield  {author} {\bibinfo {author} {\bibfnamefont {K.~A.}\ \bibnamefont {Gilmore}}, \bibinfo {author} {\bibfnamefont {M.}~\bibnamefont {Affolter}}, \bibinfo {author} {\bibfnamefont {R.}~\bibnamefont {Lewis-Swan}}, \bibinfo {author} {\bibfnamefont {D.}~\bibnamefont {Barberena}}, \bibinfo {author} {\bibfnamefont {E.}~\bibnamefont {Jordan}}, \bibinfo {author} {\bibfnamefont {A.}~\bibnamefont {Rey}},\ and\ \bibinfo {author} {\bibfnamefont {J.~J.}\ \bibnamefont {Bollinger}},\ }\bibfield  {title} {\bibinfo {title} {Quantum-enhanced sensing of displacements and electric fields with two-dimensional trapped-ion crystals},\ }\href@noop {} {\bibfield  {journal} {\bibinfo  {journal} {Science}\ }\textbf {\bibinfo {volume} {373}},\ \bibinfo {pages} {673} (\bibinfo {year} {2021})}\BibitemShut {NoStop}%
\bibitem [{\citenamefont {Deng}\ \emph {et~al.}(2024)\citenamefont {Deng}, \citenamefont {Li}, \citenamefont {Chen}, \citenamefont {Ni}, \citenamefont {Cai}, \citenamefont {Mai}, \citenamefont {Zhang}, \citenamefont {Zheng}, \citenamefont {Yu}, \citenamefont {Zou}, \citenamefont {Liu}, \citenamefont {Yan}, \citenamefont {Xu},\ and\ \citenamefont {Yu}}]{Deng2024}%
  \BibitemOpen
  \bibfield  {author} {\bibinfo {author} {\bibfnamefont {X.}~\bibnamefont {Deng}}, \bibinfo {author} {\bibfnamefont {S.}~\bibnamefont {Li}}, \bibinfo {author} {\bibfnamefont {Z.-J.}\ \bibnamefont {Chen}}, \bibinfo {author} {\bibfnamefont {Z.}~\bibnamefont {Ni}}, \bibinfo {author} {\bibfnamefont {Y.}~\bibnamefont {Cai}}, \bibinfo {author} {\bibfnamefont {J.}~\bibnamefont {Mai}}, \bibinfo {author} {\bibfnamefont {L.}~\bibnamefont {Zhang}}, \bibinfo {author} {\bibfnamefont {P.}~\bibnamefont {Zheng}}, \bibinfo {author} {\bibfnamefont {H.}~\bibnamefont {Yu}}, \bibinfo {author} {\bibfnamefont {C.-L.}\ \bibnamefont {Zou}}, \bibinfo {author} {\bibfnamefont {S.}~\bibnamefont {Liu}}, \bibinfo {author} {\bibfnamefont {F.}~\bibnamefont {Yan}}, \bibinfo {author} {\bibfnamefont {Y.}~\bibnamefont {Xu}},\ and\ \bibinfo {author} {\bibfnamefont {D.}~\bibnamefont {Yu}},\ }\bibfield  {title} {\bibinfo {title} {Quantum-enhanced metrology with large fock states},\ }\bibfield  {journal} {\bibinfo  {journal} {Nature Physics}\ }\href
  {https://doi.org/10.1038/s41567-024-02619-5} {10.1038/s41567-024-02619-5} (\bibinfo {year} {2024})\BibitemShut {NoStop}%
\bibitem [{\citenamefont {Bradley}\ \emph {et~al.}(2003)\citenamefont {Bradley}, \citenamefont {Clarke}, \citenamefont {Kinion}, \citenamefont {Rosenberg}, \citenamefont {van Bibber}, \citenamefont {Matsuki}, \citenamefont {M\"uck},\ and\ \citenamefont {Sikivie}}]{Bradley2003}%
  \BibitemOpen
  \bibfield  {author} {\bibinfo {author} {\bibfnamefont {R.}~\bibnamefont {Bradley}}, \bibinfo {author} {\bibfnamefont {J.}~\bibnamefont {Clarke}}, \bibinfo {author} {\bibfnamefont {D.}~\bibnamefont {Kinion}}, \bibinfo {author} {\bibfnamefont {L.~J.}\ \bibnamefont {Rosenberg}}, \bibinfo {author} {\bibfnamefont {K.}~\bibnamefont {van Bibber}}, \bibinfo {author} {\bibfnamefont {S.}~\bibnamefont {Matsuki}}, \bibinfo {author} {\bibfnamefont {M.}~\bibnamefont {M\"uck}},\ and\ \bibinfo {author} {\bibfnamefont {P.}~\bibnamefont {Sikivie}},\ }\bibfield  {title} {\bibinfo {title} {Microwave cavity searches for dark-matter axions},\ }\href {https://doi.org/10.1103/RevModPhys.75.777} {\bibfield  {journal} {\bibinfo  {journal} {Rev. Mod. Phys.}\ }\textbf {\bibinfo {volume} {75}},\ \bibinfo {pages} {777} (\bibinfo {year} {2003})}\BibitemShut {NoStop}%
\bibitem [{\citenamefont {Campbell}\ and\ \citenamefont {Hamilton}(2017)}]{Campbell2017}%
  \BibitemOpen
  \bibfield  {author} {\bibinfo {author} {\bibfnamefont {W.~C.}\ \bibnamefont {Campbell}}\ and\ \bibinfo {author} {\bibfnamefont {P.}~\bibnamefont {Hamilton}},\ }\bibfield  {title} {\bibinfo {title} {Rotation sensing with trapped ions*},\ }\href {https://doi.org/10.1088/1361-6455/aa5a8f} {\bibfield  {journal} {\bibinfo  {journal} {Journal of Physics B: Atomic, Molecular and Optical Physics}\ }\textbf {\bibinfo {volume} {50}},\ \bibinfo {pages} {064002} (\bibinfo {year} {2017})}\BibitemShut {NoStop}%
\bibitem [{\citenamefont {Barzanjeh}\ \emph {et~al.}(2015)\citenamefont {Barzanjeh}, \citenamefont {Guha}, \citenamefont {Weedbrook}, \citenamefont {Vitali}, \citenamefont {Shapiro},\ and\ \citenamefont {Pirandola}}]{Barzanjeh2015}%
  \BibitemOpen
  \bibfield  {author} {\bibinfo {author} {\bibfnamefont {S.}~\bibnamefont {Barzanjeh}}, \bibinfo {author} {\bibfnamefont {S.}~\bibnamefont {Guha}}, \bibinfo {author} {\bibfnamefont {C.}~\bibnamefont {Weedbrook}}, \bibinfo {author} {\bibfnamefont {D.}~\bibnamefont {Vitali}}, \bibinfo {author} {\bibfnamefont {J.~H.}\ \bibnamefont {Shapiro}},\ and\ \bibinfo {author} {\bibfnamefont {S.}~\bibnamefont {Pirandola}},\ }\bibfield  {title} {\bibinfo {title} {Microwave quantum illumination},\ }\href {https://doi.org/10.1103/PhysRevLett.114.080503} {\bibfield  {journal} {\bibinfo  {journal} {Phys. Rev. Lett.}\ }\textbf {\bibinfo {volume} {114}},\ \bibinfo {pages} {080503} (\bibinfo {year} {2015})}\BibitemShut {NoStop}%
\bibitem [{\citenamefont {Assouly}\ \emph {et~al.}(2023)\citenamefont {Assouly}, \citenamefont {Dassonneville}, \citenamefont {Peronnin}, \citenamefont {Bienfait},\ and\ \citenamefont {Huard}}]{Assouly2023}%
  \BibitemOpen
  \bibfield  {author} {\bibinfo {author} {\bibfnamefont {R.}~\bibnamefont {Assouly}}, \bibinfo {author} {\bibfnamefont {R.}~\bibnamefont {Dassonneville}}, \bibinfo {author} {\bibfnamefont {T.}~\bibnamefont {Peronnin}}, \bibinfo {author} {\bibfnamefont {A.}~\bibnamefont {Bienfait}},\ and\ \bibinfo {author} {\bibfnamefont {B.}~\bibnamefont {Huard}},\ }\bibfield  {title} {\bibinfo {title} {Quantum advantage in microwave quantum radar},\ }\href {https://doi.org/10.1038/s41567-023-02113-4} {\bibfield  {journal} {\bibinfo  {journal} {Nature Physics}\ }\textbf {\bibinfo {volume} {19}},\ \bibinfo {pages} {1418} (\bibinfo {year} {2023})}\BibitemShut {NoStop}%
\bibitem [{\citenamefont {Xiang}\ \emph {et~al.}(2017)\citenamefont {Xiang}, \citenamefont {Zhang}, \citenamefont {Jiang},\ and\ \citenamefont {Rabl}}]{Xiang2017}%
  \BibitemOpen
  \bibfield  {author} {\bibinfo {author} {\bibfnamefont {Z.-L.}\ \bibnamefont {Xiang}}, \bibinfo {author} {\bibfnamefont {M.}~\bibnamefont {Zhang}}, \bibinfo {author} {\bibfnamefont {L.}~\bibnamefont {Jiang}},\ and\ \bibinfo {author} {\bibfnamefont {P.}~\bibnamefont {Rabl}},\ }\bibfield  {title} {\bibinfo {title} {Intracity quantum communication via thermal microwave networks},\ }\href {https://doi.org/10.1103/PhysRevX.7.011035} {\bibfield  {journal} {\bibinfo  {journal} {Phys. Rev. X}\ }\textbf {\bibinfo {volume} {7}},\ \bibinfo {pages} {011035} (\bibinfo {year} {2017})}\BibitemShut {NoStop}%
\bibitem [{\citenamefont {Ball}\ and\ \citenamefont {Biercuk}(2015)}]{Ball2015}%
  \BibitemOpen
  \bibfield  {author} {\bibinfo {author} {\bibfnamefont {H.}~\bibnamefont {Ball}}\ and\ \bibinfo {author} {\bibfnamefont {M.~J.}\ \bibnamefont {Biercuk}},\ }\bibfield  {title} {\bibinfo {title} {Walsh-synthesized noise filters for quantum logic},\ }\href {https://doi.org/10.1140/epjqt/s40507-015-0022-4} {\bibfield  {journal} {\bibinfo  {journal} {EPJ Quantum Technology}\ }\textbf {\bibinfo {volume} {2}},\ \bibinfo {pages} {11} (\bibinfo {year} {2015})}\BibitemShut {NoStop}%
\bibitem [{\citenamefont {Demtröder}(2003)}]{Demtroeder2003}%
  \BibitemOpen
  \bibfield  {author} {\bibinfo {author} {\bibfnamefont {W.}~\bibnamefont {Demtröder}},\ }\href@noop {} {\emph {\bibinfo {title} {Laser Spectroscopy}}},\ \bibinfo {edition} {4th}\ ed.,\ Vol.~\bibinfo {volume} {1}\ (\bibinfo  {publisher} {Springer Berlin},\ \bibinfo {address} {Heidelberg},\ \bibinfo {year} {2003})\BibitemShut {NoStop}%
\bibitem [{\citenamefont {Fluhmann}\ and\ \citenamefont {Home}(2020)}]{Fluhmann2020}%
  \BibitemOpen
  \bibfield  {author} {\bibinfo {author} {\bibfnamefont {C.}~\bibnamefont {Fluhmann}}\ and\ \bibinfo {author} {\bibfnamefont {J.~P.}\ \bibnamefont {Home}},\ }\bibfield  {title} {\bibinfo {title} {Direct characteristic-function tomography of quantum states of the trapped-ion motional oscillator},\ }\href {https://doi.org/10.1103/PhysRevLett.125.043602} {\bibfield  {journal} {\bibinfo  {journal} {Physical Review Letters}\ }\textbf {\bibinfo {volume} {125}},\ \bibinfo {pages} {043602} (\bibinfo {year} {2020})}\BibitemShut {NoStop}%
\bibitem [{\citenamefont {Wolf}\ \emph {et~al.}(2019)\citenamefont {Wolf}, \citenamefont {Shi}, \citenamefont {Heip}, \citenamefont {Gessner}, \citenamefont {Pezzè}, \citenamefont {Smerzi}, \citenamefont {Schulte}, \citenamefont {Hammerer},\ and\ \citenamefont {Schmidt}}]{Wolf2019}%
  \BibitemOpen
  \bibfield  {author} {\bibinfo {author} {\bibfnamefont {F.}~\bibnamefont {Wolf}}, \bibinfo {author} {\bibfnamefont {C.}~\bibnamefont {Shi}}, \bibinfo {author} {\bibfnamefont {J.~C.}\ \bibnamefont {Heip}}, \bibinfo {author} {\bibfnamefont {M.}~\bibnamefont {Gessner}}, \bibinfo {author} {\bibfnamefont {L.}~\bibnamefont {Pezzè}}, \bibinfo {author} {\bibfnamefont {A.}~\bibnamefont {Smerzi}}, \bibinfo {author} {\bibfnamefont {M.}~\bibnamefont {Schulte}}, \bibinfo {author} {\bibfnamefont {K.}~\bibnamefont {Hammerer}},\ and\ \bibinfo {author} {\bibfnamefont {P.~O.}\ \bibnamefont {Schmidt}},\ }\bibfield  {title} {\bibinfo {title} {Motional fock states for quantum-enhanced amplitude and phase measurements with trapped ions},\ }\href {https://doi.org/10.1038/s41467-019-10576-4} {\bibfield  {journal} {\bibinfo  {journal} {Nature Communications}\ }\textbf {\bibinfo {volume} {10}},\ \bibinfo {pages} {2929} (\bibinfo {year} {2019})}\BibitemShut {NoStop}%
\bibitem [{\citenamefont {Wu}\ \emph {et~al.}(2025{\natexlab{b}})\citenamefont {Wu}, \citenamefont {Ho}, \citenamefont {Mitts}, \citenamefont {Rabinowitz},\ and\ \citenamefont {Hudson}}]{WuH2025}%
  \BibitemOpen
  \bibfield  {author} {\bibinfo {author} {\bibfnamefont {H.}~\bibnamefont {Wu}}, \bibinfo {author} {\bibfnamefont {C.~Z.~C.}\ \bibnamefont {Ho}}, \bibinfo {author} {\bibfnamefont {G.~D.}\ \bibnamefont {Mitts}}, \bibinfo {author} {\bibfnamefont {J.~A.}\ \bibnamefont {Rabinowitz}},\ and\ \bibinfo {author} {\bibfnamefont {E.~R.}\ \bibnamefont {Hudson}},\ }\href {https://arxiv.org/abs/2506.03010} {\bibinfo {title} {Floquet-engineered decoherence-resilient protocols for wideband sensing beyond the linear standard quantum limit}} (\bibinfo {year} {2025}{\natexlab{b}}),\ \Eprint {https://arxiv.org/abs/2506.03010} {arXiv:2506.03010 [physics.atom-ph]} \BibitemShut {NoStop}%
\bibitem [{\citenamefont {McCormick}\ \emph {et~al.}(2019)\citenamefont {McCormick}, \citenamefont {Keller}, \citenamefont {Burd}, \citenamefont {Wineland}, \citenamefont {Wilson},\ and\ \citenamefont {Leibfried}}]{McCormick2019}%
  \BibitemOpen
  \bibfield  {author} {\bibinfo {author} {\bibfnamefont {K.~C.}\ \bibnamefont {McCormick}}, \bibinfo {author} {\bibfnamefont {J.}~\bibnamefont {Keller}}, \bibinfo {author} {\bibfnamefont {S.~C.}\ \bibnamefont {Burd}}, \bibinfo {author} {\bibfnamefont {D.~J.}\ \bibnamefont {Wineland}}, \bibinfo {author} {\bibfnamefont {A.~C.}\ \bibnamefont {Wilson}},\ and\ \bibinfo {author} {\bibfnamefont {D.}~\bibnamefont {Leibfried}},\ }\bibfield  {title} {\bibinfo {title} {Quantum-enhanced sensing of a single-ion mechanical oscillator},\ }\href {https://doi.org/10.1038/s41586-019-1421-y} {\bibfield  {journal} {\bibinfo  {journal} {Nature}\ }\textbf {\bibinfo {volume} {572}},\ \bibinfo {pages} {86} (\bibinfo {year} {2019})}\BibitemShut {NoStop}%
\bibitem [{\citenamefont {Kaubruegger}\ \emph {et~al.}(2021)\citenamefont {Kaubruegger}, \citenamefont {Vasilyev}, \citenamefont {Schulte}, \citenamefont {Hammerer},\ and\ \citenamefont {Zoller}}]{Kaubruegger2021}%
  \BibitemOpen
  \bibfield  {author} {\bibinfo {author} {\bibfnamefont {R.}~\bibnamefont {Kaubruegger}}, \bibinfo {author} {\bibfnamefont {D.~V.}\ \bibnamefont {Vasilyev}}, \bibinfo {author} {\bibfnamefont {M.}~\bibnamefont {Schulte}}, \bibinfo {author} {\bibfnamefont {K.}~\bibnamefont {Hammerer}},\ and\ \bibinfo {author} {\bibfnamefont {P.}~\bibnamefont {Zoller}},\ }\bibfield  {title} {\bibinfo {title} {Quantum variational optimization of ramsey interferometry and atomic clocks},\ }\href {https://doi.org/10.1103/PhysRevX.11.041045} {\bibfield  {journal} {\bibinfo  {journal} {Phys. Rev. X}\ }\textbf {\bibinfo {volume} {11}},\ \bibinfo {pages} {041045} (\bibinfo {year} {2021})}\BibitemShut {NoStop}%
\bibitem [{\citenamefont {Marciniak}\ \emph {et~al.}(2022)\citenamefont {Marciniak}, \citenamefont {Feldker}, \citenamefont {Pogorelov}, \citenamefont {Kaubruegger}, \citenamefont {Vasilyev}, \citenamefont {van Bijnen}, \citenamefont {Schindler}, \citenamefont {Zoller}, \citenamefont {Blatt},\ and\ \citenamefont {Monz}}]{Marciniak2022}%
  \BibitemOpen
  \bibfield  {author} {\bibinfo {author} {\bibfnamefont {C.~D.}\ \bibnamefont {Marciniak}}, \bibinfo {author} {\bibfnamefont {T.}~\bibnamefont {Feldker}}, \bibinfo {author} {\bibfnamefont {I.}~\bibnamefont {Pogorelov}}, \bibinfo {author} {\bibfnamefont {R.}~\bibnamefont {Kaubruegger}}, \bibinfo {author} {\bibfnamefont {D.~V.}\ \bibnamefont {Vasilyev}}, \bibinfo {author} {\bibfnamefont {R.}~\bibnamefont {van Bijnen}}, \bibinfo {author} {\bibfnamefont {P.}~\bibnamefont {Schindler}}, \bibinfo {author} {\bibfnamefont {P.}~\bibnamefont {Zoller}}, \bibinfo {author} {\bibfnamefont {R.}~\bibnamefont {Blatt}},\ and\ \bibinfo {author} {\bibfnamefont {T.}~\bibnamefont {Monz}},\ }\bibfield  {title} {\bibinfo {title} {Optimal metrology with programmable quantum sensors},\ }\href {https://doi.org/10.1038/s41586-022-04435-4} {\bibfield  {journal} {\bibinfo  {journal} {Nature}\ }\textbf {\bibinfo {volume} {603}},\ \bibinfo {pages} {604} (\bibinfo {year} {2022})}\BibitemShut {NoStop}%
\bibitem [{\citenamefont {Meyer}\ \emph {et~al.}(2021)\citenamefont {Meyer}, \citenamefont {Borregaard},\ and\ \citenamefont {Eisert}}]{Meyer2021}%
  \BibitemOpen
  \bibfield  {author} {\bibinfo {author} {\bibfnamefont {J.~J.}\ \bibnamefont {Meyer}}, \bibinfo {author} {\bibfnamefont {J.}~\bibnamefont {Borregaard}},\ and\ \bibinfo {author} {\bibfnamefont {J.}~\bibnamefont {Eisert}},\ }\bibfield  {title} {\bibinfo {title} {A variational toolbox for quantum multi-parameter estimation},\ }\href {https://doi.org/10.1038/s41534-021-00425-y} {\bibfield  {journal} {\bibinfo  {journal} {npj Quantum Information}\ }\textbf {\bibinfo {volume} {7}},\ \bibinfo {pages} {89} (\bibinfo {year} {2021})}\BibitemShut {NoStop}%
\bibitem [{\citenamefont {Schlawin}\ \emph {et~al.}(2022)\citenamefont {Schlawin}, \citenamefont {Kennes},\ and\ \citenamefont {Sentef}}]{Schlawin2022}%
  \BibitemOpen
  \bibfield  {author} {\bibinfo {author} {\bibfnamefont {F.}~\bibnamefont {Schlawin}}, \bibinfo {author} {\bibfnamefont {D.~M.}\ \bibnamefont {Kennes}},\ and\ \bibinfo {author} {\bibfnamefont {M.~A.}\ \bibnamefont {Sentef}},\ }\bibfield  {title} {\bibinfo {title} {Cavity quantum materials},\ }\href {https://doi.org/10.1063/5.0083825} {\bibfield  {journal} {\bibinfo  {journal} {Applied Physics Reviews}\ }\textbf {\bibinfo {volume} {9}},\ \bibinfo {pages} {011312} (\bibinfo {year} {2022})},\ \Eprint {https://arxiv.org/abs/https://pubs.aip.org/aip/apr/article-pdf/doi/10.1063/5.0083825/19819541/011312\_1\_online.pdf} {https://pubs.aip.org/aip/apr/article-pdf/doi/10.1063/5.0083825/19819541/011312\_1\_online.pdf} \BibitemShut {NoStop}%
\bibitem [{\citenamefont {Krantz}\ \emph {et~al.}(2019)\citenamefont {Krantz}, \citenamefont {Kjaergaard}, \citenamefont {Yan}, \citenamefont {Orlando}, \citenamefont {Gustavsson},\ and\ \citenamefont {Oliver}}]{Krantz2019}%
  \BibitemOpen
  \bibfield  {author} {\bibinfo {author} {\bibfnamefont {P.}~\bibnamefont {Krantz}}, \bibinfo {author} {\bibfnamefont {M.}~\bibnamefont {Kjaergaard}}, \bibinfo {author} {\bibfnamefont {F.}~\bibnamefont {Yan}}, \bibinfo {author} {\bibfnamefont {T.~P.}\ \bibnamefont {Orlando}}, \bibinfo {author} {\bibfnamefont {S.}~\bibnamefont {Gustavsson}},\ and\ \bibinfo {author} {\bibfnamefont {W.~D.}\ \bibnamefont {Oliver}},\ }\bibfield  {title} {\bibinfo {title} {A quantum engineer's guide to superconducting qubits},\ }\href {https://doi.org/10.1063/1.5089550} {\bibfield  {journal} {\bibinfo  {journal} {Applied Physics Reviews}\ }\textbf {\bibinfo {volume} {6}},\ \bibinfo {pages} {021318} (\bibinfo {year} {2019})},\ \Eprint {https://arxiv.org/abs/https://pubs.aip.org/aip/apr/article-pdf/doi/10.1063/1.5089550/16667201/021318\_1\_online.pdf} {https://pubs.aip.org/aip/apr/article-pdf/doi/10.1063/1.5089550/16667201/021318\_1\_online.pdf} \BibitemShut {NoStop}%
\bibitem [{\citenamefont {Wang}\ \emph {et~al.}(2022)\citenamefont {Wang}, \citenamefont {Liu}, \citenamefont {Schloss}, \citenamefont {Alsid}, \citenamefont {Braje},\ and\ \citenamefont {Cappellaro}}]{Wang2022}%
  \BibitemOpen
  \bibfield  {author} {\bibinfo {author} {\bibfnamefont {G.}~\bibnamefont {Wang}}, \bibinfo {author} {\bibfnamefont {Y.-X.}\ \bibnamefont {Liu}}, \bibinfo {author} {\bibfnamefont {J.~M.}\ \bibnamefont {Schloss}}, \bibinfo {author} {\bibfnamefont {S.~T.}\ \bibnamefont {Alsid}}, \bibinfo {author} {\bibfnamefont {D.~A.}\ \bibnamefont {Braje}},\ and\ \bibinfo {author} {\bibfnamefont {P.}~\bibnamefont {Cappellaro}},\ }\bibfield  {title} {\bibinfo {title} {Sensing of arbitrary-frequency fields using a quantum mixer},\ }\href {https://doi.org/10.1103/PhysRevX.12.021061} {\bibfield  {journal} {\bibinfo  {journal} {Phys. Rev. X}\ }\textbf {\bibinfo {volume} {12}},\ \bibinfo {pages} {021061} (\bibinfo {year} {2022})}\BibitemShut {NoStop}%
\bibitem [{\citenamefont {Demkowicz-Dobrzański}\ \emph {et~al.}(2015)\citenamefont {Demkowicz-Dobrzański}, \citenamefont {Jarzyna},\ and\ \citenamefont {Kołodyński}}]{wolf2015}%
  \BibitemOpen
  \bibfield  {author} {\bibinfo {author} {\bibfnamefont {R.}~\bibnamefont {Demkowicz-Dobrzański}}, \bibinfo {author} {\bibfnamefont {M.}~\bibnamefont {Jarzyna}},\ and\ \bibinfo {author} {\bibfnamefont {J.}~\bibnamefont {Kołodyński}},\ }\bibfield  {title} {\bibinfo {title} {Chapter four - quantum limits in optical interferometry}\ }(\bibinfo  {publisher} {Elsevier},\ \bibinfo {year} {2015})\ pp.\ \bibinfo {pages} {345--435}\BibitemShut {NoStop}%
\end{thebibliography}%
